\documentclass[aps,prl,superscriptaddress,twocolumn,floatfix,nofootinbib]{revtex4-2}

\usepackage{amsmath,amssymb,bm,slashed}
\usepackage{graphicx}
\usepackage{multirow}
\usepackage{longtable}
\usepackage{orcidlink} 
\usepackage[normalem]{ulem}
\usepackage{xcolor}

\hypersetup{colorlinks=true,linkcolor=blue,citecolor=blue,urlcolor=blue}

\newcommand{\runin}[1]{\noindent\textit{#1}---\ignorespaces}

\newcommand{\sxu}{\affiliation{College of Physics and Electronic Engineering, Shanxi University, Taiyuan 030006, China}}
\newcommand{\ucas}{\affiliation{School of Physical Sciences, University of Chinese Academy of Sciences, Beijing 100049, China}}
\newcommand{\scnt}{\affiliation{Southern Center for Nuclear-Science Theory, Institute of Modern Physics, Chinese Academy of Sciences, Huizhou 516000, China}}

\begin{document}
	
	\preprint{}
	
	\title{
		A General Manifestly Covariant Construction of Partial-Wave Amplitudes in the Rarita-Schwinger Formalism
	}
	
	\author{Hao-Jie Jing\orcidlink{0000-0002-3651-5722}}
	\email[Corresponding author: ]{jinghaojie@sxu.edu.cn}
	\sxu
	\author{Shu-Ming Wu\orcidlink{0000-0002-8658-9789}}
	\email{wushuming@ucas.ac.cn}
	\ucas
	\author{Jia-Jun Wu\orcidlink{0000-0003-4583-7691}}
	\email[Corresponding author: ]{wujiajun@ucas.ac.cn}
	\ucas\scnt

	\date{\today}
	
	\begin{abstract}
			We propose a new set of functions, named Relative Spin Wave Functions (RSWFs), defined via covariant pure boosts to a common rest frame. 
			RSWFs separate spin transport from orbital-spin coupling while preserving the conventional partial-wave angular structure in a manifestly covariant form. 
			Within the Rarita-Schwinger formalism, this approach generates explicit, covariant on-shell partial-wave amplitudes for two-body decays of massive particles of arbitrary spin, with definite orbital angular momentum and total final-state spin. 
			This general framework provides a foundational Lorentz basis for hadronic reaction studies and $LS$-organized operators in theoretical investigations, such as Lattice QCD and QCD sum rules.
	\end{abstract}
	\maketitle

	\runin{\bf Introduction}
	Partial-wave analysis (PWA) is a widely used method in high-energy particle physics experiments for extracting the spin-parity quantum numbers of particles, and thereby for studying the classification of these particles and the interaction patterns among them. 
	PWA has played a crucial role both in the establishment of the Standard Model of particle physics in the last century, where it provided the spin-parity assignments of the observed hadron resonances~\cite{Workman:2022ynf}, and in the ongoing classification of exotic hadronic states and the study of their decay properties. 
	For a systematic introduction to the theoretical methods of PWA and a comparison of different PWA schemes, we refer the reader to a recent review~\cite{Huang:2026egv} and the references therein.
	
	In a previous work~\cite{Jing:2024mag}, the partial-wave amplitudes of a two-body decay process ($1\to2+3$) with orbital-spin quantum numbers $L$-$S$ were classified into two types, $\mathcal{H}$-type and $\mathcal{C}$-type, according to the angular distributions they produce.
	The existing PWA schemes can all be classified into these two types. 
	Examples of $\mathcal{H}$-type amplitudes include the $LS$ amplitudes obtained from the Jacob--Wick helicity formalism~\cite{Jacob:1959at}, and the covariant canonical-spinor amplitudes of Ref.~\cite{Huang:2026egv}.
	In contrast, the partial-wave amplitudes given in Refs.~\cite{Chung:1993da,Chung:1997jn,Zou:2002ar} are of $\mathcal{C}$-type.
	The specific forms of the two types of partial-wave amplitudes in the center-of-mass (c.m.) frame read, respectively,
	\begin{equation} 
		\mathcal{H}_{L, S}^*=~\underbrace{\Gamma\left(k_1, p_2^*, p_3^* ; L, S\right)}_{\text{$LS$ coupling structure}} \,\cdot\, \underbrace{\bar{u}\left(k_1;s_1\right) u\left(k_2;s_2\right) u\left(k_3;s_3\right)}_{\text{pure-spin part}},  
		\label{eq:pwamp-H-cmf}
	\end{equation}
	\begin{equation}
		\mathcal{C}_{L, S}^*=~\underbrace{\Gamma\left(k_1, p_2^*, p_3^* ; L, S\right)}_{\text{$LS$ coupling structure}} \,\cdot\, \underbrace{\bar{u}\left(k_1;s_1\right) u\left(p_2^*; s_2\right) u\left(p_3^*; s_3\right)}_{\text{general-spin part}},
		\label{eq:pwamp-C-cmf} 
	\end{equation}
	where $k_i=(m_i,\bm{0})$ $(i=1,2,3)$ are the four-momenta of the particles in their own rest frames, and $p_2^*, p_3^*$ are the four-momenta of the final particles in the c.m.\ frame; $u\left(p_i ; s_i\right)$ collectively denotes the external spin wave function of particle $i$ with the canonical spin projection label suppressed for brevity. %
	Accordingly, the pure-spin part of the $\mathcal{H}$-type amplitude carries no angular dependence. 
	In Ref.~\cite{Huang:2026egv}, the $\mathcal{H}$-type and $\mathcal{C}$-type partial-wave amplitudes are referred to as the partial-wave amplitudes in the pure-spin and general-spin schemes, respectively. 
	The nomenclature of the spin-wave-function parts in Eqs.~\eqref{eq:pwamp-H-cmf}--\eqref{eq:pwamp-C-cmf} follows that reference.

	Since the angular distribution of the $\mathcal{H}$-type amplitude is entirely determined by the coupling structure tensor $\Gamma\left(k_1, p_2^*, p_3^* ; L, S\right)$, one can show that it coincides with that of the conventional partial-wave amplitude (see Ref.~\cite{Huang:2026egv} for a detailed proof), i.e., the c.m.\ angular distribution is governed by the spherical harmonic $\mathcal{Y}_{L\ell}(\hat{\bm{q}}^*)$, where $\bm{q}^*$ is the spatial component of the relative four-momentum $q^*=(p_2^*-p_3^*)/2$ in the c.m.\ frame. 
	In contrast, the general-spin part of a $\mathcal C$-type amplitude can introduce additional angular dependence, so its orbital label need not coincide with the relative orbital angular momentum in the two-body c.m.\ frame.
	The two constructions have the same angular structure in the
	nonrelativistic limit; this agreement can also hold at finite
	velocities when the complete contraction eliminates the
	boost-dependent terms.
	In particular, in the non-relativistic approximation ($p_2^*,p_3^*\to k_2,k_3$),
	\begin{equation}
		u\left(p_2^*; s_2\right) u\left(p_3^*; s_3\right)~\mapsto~u\left(k_2; s_2\right) u\left(k_3; s_3\right),
	\end{equation}
	and, for every fixed pair of quantum numbers $L$-$S$, the $\mathcal{H}$-type and $\mathcal{C}$-type amplitudes have the same angular distribution, coinciding with that of the corresponding conventional partial wave.	
	Since the $\mathcal{C}$-type amplitude is manifestly covariant, its form in an arbitrary reference frame is obtained by simply replacing the c.m.\ momentum triplet $(k_1,p_2^*,p_3^*)$ with the triplet $(p_1,p_2,p_3)$ in that frame, see Eq.~\eqref{eq:pwamp-C-abf}. 
	The $\mathcal{H}$-type amplitude defined in Eq.~\eqref{eq:pwamp-H-cmf}, however, is not manifestly covariant: to evaluate it in an arbitrary frame, an additional alignment rotation must be applied to Eq.~\eqref{eq:pwamp-H-cmf}; for details see Ref.~\cite{ChouShirokov1958}.
	The $\mathcal{C}$-type construction is manifestly covariant but can mix conventional partial waves through its momentum-dependent spin factors. 
	The c.m.-frame $\mathcal{H}$-type construction preserves the conventional $L$-$S$ decomposition, but evaluating it in a common frame requires momentum-dependent boosts and spin
	rotations~\cite{Jing:2024mag}.
	When vertices defined in different two-body rest frames are embedded in loop integrals or relativistic scattering equations, these transformations enter the integrands and 	scattering kernels, complicating their construction and evaluation.
	A manifestly covariant $\mathcal{H}$-type formulation in Lorentz-vector and Dirac-spinor indices is therefore particularly valuable: it preserves the conventional partial-wave interpretation while expressing spin transport through covariant tensor contractions, providing a natural starting point for relativistic dynamical calculations.

	Manifestly covariant $\mathcal{H}$-type amplitudes have recently been constructed using covariant canonical spinors~\cite{Huang:2026egv}. 
	However, that work does not provide a general set of explicit formulas in Lorentz-vector and Dirac-spinor indices for arbitrary massive spins.
	This leaves the last piece of the puzzle unresolved: how to systematically construct manifestly covariant $\mathcal{H}$-type amplitudes in the Rarita--Schwinger (RS) formalism~\cite{Rarita:1941mf}.

	In this Letter, we address this question by introducing	relative spin wave functions (RSWFs), which separate covariant spin transport from orbital-spin coupling and yield explicit $\mathcal{H}$-type amplitudes directly in the RS formalism.

	\runin{\bf RSWFs}
	The relative spin wave function $u\left(p_2;s_2|p_1\right)$ of particle 2 with respect to particle 1 (referred to as the anchor particle) is defined by the following two conditions:
	\begin{itemize}
		\item \emph{Manifest covariance}---under Lorentz transformations, the relative spin wave function $u\left(p_2;s_2|p_1\right)$ transforms in exactly the same way as the spin wave function $u\left(p_2;s_2\right)$.
		\item \emph{Anchor property}---in the frame where the spin wave function of the anchor particle is $u\left(k_1;s_1\right)$ and the corresponding four momentum of particle 2 is $p_2^*$, 	the relative spin wave function satisfies $u\left(p_2^*;s_2|k_1\right)\equiv f(p_1\cdot p_2)\, u\left(k_2;s_2\right)$. 
		The scalar factor $f(p_1\cdot p_2)$ is independent
		of the momentum direction in the anchor rest frame and
		therefore does not alter the partial-wave angular structure.
	\end{itemize}
	It is then easy to show that the general form of the relative spin wave function $u\left(p_2;s_2|p_1\right)$ reads
	\begin{equation}
		u\left(p_2;s_2|p_1\right)=~F_{s_2}(v_1\cdot v_2)~B(v_1,v_2)~u\left(p_2;s_2\right),
		\label{eq:def-relative-swf}
	\end{equation}
	where $v_i=p_i/m_i~(i=1,2)$ are the four-velocities of the particles, $B(v',v)$ denotes the pure boost element that transforms the future-directed timelike four-velocity $v$ into $v'$ and $F_{s_2}(x)$ is an arbitrary normalization function satisfying $F_{s_2}(1)=1$.
	In particular, for a vector particle, substituting the manifestly covariant four-vector form of the pure boost element in the fundamental representation of the Lorentz group (for the explicit expression, see Refs.~\cite{10.1063/1.523355,10.1119/1.12748}) into Eq.~\eqref{eq:def-relative-swf}, one obtains the relative polarization vector
	\begin{align}
		\varepsilon_\mu(p_2|p_1)=&\left(v^2\delta_\mu^{\nu}-2v_\mu v^\nu\right)\varepsilon_\nu(p_2)
		=\left(v^2\delta_\mu^{\nu}-v_\mu v_1^\nu\right)\varepsilon_\nu(p_2)\notag\\
		\equiv&~\Pi_\mu^\nu(p_1,p_2)\,\varepsilon_\nu(p_2),
	\end{align}
	where $v\equiv(v_1+v_2)/2$ and we have chosen the normalization function $F_1(v_1\cdot v_2)=v^2$. 
	The second equality follows from the transversality condition $v_2\cdot\varepsilon(p_2)=0$. 
	One can verify that the relative polarization vector satisfies $\varepsilon(p_1|p_1)=\varepsilon(p_1)$ and $p_1\cdot\varepsilon(p_2|p_1)=0$.
	For spin 1, the normalized transport kernel
	$\Pi_\mu^\nu(p_1,p_2)/v^2$, acting on transverse polarization
	vectors, also appears in the fixed-axis polarization construction
	in Ref.~\cite{Gross:2008}.
	
	For a spin-$1/2$ fermion, substituting the manifestly covariant four-vector form of the pure boost element in the Dirac spinor representation (for the explicit expression, see the appendix of Ref.~\cite{10.1063/1.523355}) into Eq.~\eqref{eq:def-relative-swf}, one obtains the relative Dirac spinors
	\begin{align}
		u(p_2|p_1)=&\frac{1+\slashed{v}_1\slashed{v}_2}{2}~u(p_2)
		=\frac{1+\slashed{v}_1}{2}~u(p_2)\notag\\
		\equiv&~ \Gamma_+(p_1)\,u(p_2),
	\end{align}
	\begin{align}
		v(p_2|p_1)=&\frac{1+\slashed{v}_1\slashed{v}_2}{2}~v(p_2)
		=\frac{1-\slashed{v}_1}{2}~v(p_2)\notag\\
		\equiv&~ \Gamma_-(p_1)\,v(p_2),
	\end{align}
	where we have chosen the normalization function $F_{\frac12}(v_1\cdot v_2)=\sqrt{v^2}$. 
	The second equalities in the above two equations follow from the on-shell conditions $\slashed{v}_2 u(p_2)=u(p_2)$ and $\slashed{v}_2 v(p_2)=-v(p_2)$, respectively. 
	Evidently, the relative Dirac spinors satisfy $u(p_1|p_1)=u(p_1)$, $v(p_1|p_1)=v(p_1)$, and $\Gamma_+(p_1)v(p_2|p_1)=\Gamma_-(p_1)u(p_2|p_1)=0$.
	
	In general, in the RS form, the relative spin wave function of a particle of spin $s_2$ is defined as
	\begin{align}
		&\varepsilon_{\bm{\mu}}(p_2|p_1)=\Pi_{\bm{\mu}}^{\bm{\nu}}(p_1,p_2)~\varepsilon_{\bm{\nu}}(p_2),
		\label{eq:relative-swf-vareps}\\
		&u_{\bm{\mu}}(p_2|p_1)=\Gamma_+(p_1)\, \Pi_{\bm{\mu}}^{\bm{\nu}}(p_1,p_2)~u_{\bm{\nu}}(p_2),
		\label{eq:relative-swf-u}\\
		&v_{\bm{\mu}}(p_2|p_1)=\Gamma_-(p_1)\, \Pi_{\bm{\mu}}^{\bm{\nu}}(p_1,p_2)~v_{\bm{\nu}}(p_2),
		\label{eq:relative-swf-v}
	\end{align}
	where, for brevity, the bold Greek letters $\bm{\mu}$ and $\bm{\nu}$ denote the Lorentz index sets $\mu_1\mu_2\cdots$ and $\nu_1\nu_2\cdots$, respectively. 
	The numbers of indices in these index sets are $s_2$ and $s_2-1/2$ for bosons and fermions, respectively. 
	The object $\Pi_{\bm{\mu}}^{\bm{\nu}}(p_1,p_2)=\Pi_{\mu_1}^{\nu_1}(p_1,p_2)\Pi_{\mu_2}^{\nu_2}(p_1,p_2)\cdots$ denotes the multiple tensor product of $\Pi_\mu^\nu(p_1,p_2)$. 
	The normalization function $F_{s_2}(v_1\cdot v_2)=(v^2)^{s_2}$ has been absorbed into the definitions of $\Pi$ and $\Gamma_\pm$. 
	One can verify that the RSWFs satisfy the kinematic constraints
	of a massive spin-$s_2$ wave function referred to the anchor
	velocity $v_1$, and that they reduce to the corresponding
	ordinary spin wave functions when $v_2=v_1$.
	Here the kinematic constraints refer to the transversality, symmetry, tracelessness,
	$\gamma$-tracelessness, and Dirac constraints, as appropriate
	for spin $s_2$, with all momentum-dependent constraints
	referred to $v_1$.
	
	The spin-$1/2$ particle wave function above coincides with the on-shell heavy-particle spinor in Ref.~\cite{Aoude:2020HPET}; the explicit higher-spin examples in that work belong to the same family with different scalar normalizations once the spin bases are aligned.
	Here Lorentz covariance and the anchor property define RSWFs independently of an effective-theory expansion.
	These wave functions describe spin relative to an anchor: in its rest frame, the spin structure is independent of the relative momentum direction, and ordinary wave functions are recovered at coincident velocities.
	This characterization unifies these realizations and supplies the spin building blocks for manifestly covariant pure-$LS$ amplitudes for arbitrary massive spins.

	With RSWFs, the manifestly covariant form of the $\mathcal{H}$-type partial-wave amplitude is defined as
	\begin{align}
		\mathcal{H}_{L, S}\equiv&~\Gamma\left(p_1, p_2, p_3; L, S\right)
		\notag\\&\times
		\bar{u}\left(p_1;s_1\right) u\left(p_2;s_2|p_1\right) u\left(p_3;s_3|p_1\right).
		\label{eq:pwamp-H-abf}
	\end{align}
	When returning to the rest frame of the initial particle, Eq.~\eqref{eq:pwamp-H-abf} reduces to Eq.~\eqref{eq:pwamp-H-cmf}, up to an angle-independent normalization factor $F_{s_2}(v_1\cdot v_2)F_{s_3}(v_1\cdot v_3)$.
	
	For comparison, the manifestly covariant form of the $\mathcal{C}$-type partial-wave amplitude is defined as
	\begin{align}
		\mathcal{C}_{L, S}\equiv\Gamma\left(p_1, p_2, p_3; L, S\right)\,\times\,
		\bar{u}\left(p_1;s_1\right) u\left(p_2;s_2\right) u\left(p_3;s_3\right).
		\label{eq:pwamp-C-abf}
	\end{align}
	Comparing Eqs.~\eqref{eq:pwamp-H-abf} and \eqref{eq:pwamp-C-abf}, one sees that the only difference between the $\mathcal{H}$-type and $\mathcal{C}$-type partial-wave amplitudes lies in whether the spin wave functions of the external particles are the ordinary or the relative ones.

	\runin{\bf Coupling structure tensor}
	For the general causal-field form of the coupling structure tensor $\Gamma\left(p_1, p_2, p_3; L, S\right)$, see Ref.~\cite{Jing:2023rwz}. 
	In the RS form, in particular, it reads
	\begin{equation}
		\Gamma_{\bm{\mu}}^{\bm{\nu}\bm{\rho}}\left(p_1, p_2, p_3; L, S\right)=P_{s_1\bm{\mu}}^{L\bm{\xi};S\bm{\sigma}}(p_1)\,P_{S\bm{\sigma}}^{s_2\bm{\nu};s_3\bm{\rho}}(p_1)\,q_{\bm{\xi}},
		\label{eq:Gamma-PPq}
	\end{equation}
	where $q_{\bm{\xi}}=q_{\xi_1}\cdots q_{\xi_L}$ is the multiple tensor product of the relative four-momentum $q$. 
	The tensor $P_{S\bm{\sigma}}^{s_2\bm{\nu};s_3\bm{\rho}}(p_1)$ couples the two final spins to $S$, while $P_{s_1\bm{\mu}}^{L\bm{\xi};S\bm{\sigma}}(p_1)$ couples
	$L$ and $S$ to $s_1$.
	These tensors are covariant counterparts of the SU(2) Clebsch--Gordan coefficients (CGCs) in the RS formalism.
	In general, an angular momentum coupling tensor $P_{s_1\bm{\mu}}^{s_2\bm{\nu};s_3\bm{\rho}}(p_1)$ can be constructed from the spin projection tensors $P^{(s)}_{\bm{\mu};\bm{\nu}}(p_1)$, the Dirac $\gamma$-matrices, the totally antisymmetric tensor $\epsilon_{\mu\nu\rho\sigma}$, and the four-velocity $v_1$. 
	In the following, we provide the explicit expressions of the angular momentum coupling tensors in three cases, classified according to which of the external particles carry half-integer spins; within each case, the value of $\delta (\equiv s_1+s_2+s_3\mod{2})$ distinguishes two possibilities.

	\begin{itemize}
		\item Case I 
		
		$s_2,s_3$ are both integer spins (in this case $s_1$ is necessarily an integer spin).
		For $\delta=0$ the coupling structure tensor reads
		\begin{align}
			P_{s_1\bm{\mu}}^{s_2\bm{\nu};s_3\bm{\rho}}(p_1)=&~P^{(s_1)}_{\bm{\mu};\bm{\alpha}\bm{\gamma}}(p_1)\,P^{(s_2)\bm{\nu};\bm{\alpha}\bm{\beta}}(p_1)\notag\\
			&\times P^{(s_3)\bm{\gamma}\bm{\beta}';\bm{\rho}}(p_1)\,g_{\bm{\beta},\bm{\beta}'},
			\label{eq:AMCT-1A}
		\end{align}
		where $g_{\bm{\beta},\bm{\beta}'}=g_{\beta_1\beta_1'}g_{\beta_2\beta_2'}\cdots$ is the multiple tensor product of the metric $g_{\beta\beta'}$, and for the explicit form of the integer-spin projection tensor $P^{(n)}_{\bm{\mu};\bm{\nu}}(p_1)~(n\in\mathbb{N})$ see Eq.~(8) of Ref.~\cite{Behrends:1957rup} or Eq.~(7) of Ref.~\cite{Huang:2002zt}. 
		It should be noted that, although we have referred to the coupling structure tensors as the covariant versions of the SU(2) CGCs, the normalization condition of the coupling tensors given in this Letter differs from that of the conventional CGCs: in general, $P_{s_1\bm{\mu}}^{s_2\bm{\nu};s_3\bm{\rho}}(p_1)P^{s_1\bm{\mu}}_{s_2\bm{\nu};s_3\bm{\rho}}(p_1)\neq 2s_1+1$. 
		To reproduce the conventional CGC normalization, one would only need to multiply each coupling tensor by a constant factor. 
		Such factors, however, are irrelevant for the construction of the covariant on-shell partial-wave amplitudes, and for simplicity they are not considered in this work.
		For $\delta=1$ it reads
		\begin{align}
			P_{s_1\bm{\mu}}^{s_2\bm{\nu};s_3\bm{\rho}}(p_1)=&~iP^{(s_1)}_{\bm{\mu};\bm{\alpha}\bm{\gamma}\lambda}(p_1)\,P^{(s_2)\bm{\nu};\bm{\alpha}\bm{\beta}\kappa}(p_1)\notag\\
			&\times P^{(s_3)\bm{\gamma}\bm{\beta}'\tau;\bm{\rho}}(p_1)\,g_{\bm{\beta},\bm{\beta}'}\,\tilde{\epsilon}^{\lambda}_{\kappa\tau}(p_1),
			\label{eq:AMCT-1B}
		\end{align}
		where $\tilde{\epsilon}^{\lambda}_{\kappa\tau}(p_1)=\epsilon_{\kappa\tau\alpha\beta}\,g^{\alpha\lambda}\,v_1^\beta$.
		The numbers of indices in the index sets $\bm{\mu},\bm{\nu},\bm{\rho}$ and $\bm{\alpha},\bm{\beta}^{(\prime)},\bm{\gamma}$ in Eqs.~\eqref{eq:AMCT-1A} and \eqref{eq:AMCT-1B} are $s_1,s_2,s_3$ and $a_\mathrm{I},b_\mathrm{I},c_\mathrm{I}$, respectively. 
		The triplet $(a_\mathrm{I},b_\mathrm{I},c_\mathrm{I})$ characterizing the numbers of indices reads
		\begin{align}
			a_\mathrm{I} & =\frac{s_1+s_2-s_3-\delta}{2}, 
			\label{eq:aI}\\
			b_\mathrm{I} & =\frac{s_2+s_3-s_1-\delta}{2}, 
			\label{eq:bI}\\
			c_\mathrm{I} & =\frac{s_1+s_3-s_2-\delta}{2}.
			\label{eq:cI}
		\end{align}

		\item Case II
		
		$s_2,s_3$ are both half-integer spins (in this case $s_1$ is necessarily an integer spin).
		For $\delta=0$ the coupling structure tensor reads
		\begin{align}
			P_{s_1\bm{\mu}}^{s_2\bm{\nu};s_3\bm{\rho}}(p_1)=&~P^{(s_2)\bm{\nu};\bm{\alpha}\bm{\beta}}(p_1)\,\gamma_5\gamma^{\lambda}\,P^{(s_3)\bm{\gamma}\bm{\beta}';\bm{\rho}}(p_1)\notag\\&\times P^{(s_1)}_{\bm{\mu};\bm{\alpha}\bm{\gamma}\lambda}(p_1)\,g_{\bm{\beta},\bm{\beta}'}.
			\label{eq:AMCT-2A}
		\end{align}
		For $\delta=1$ it reads
		\begin{align}
			P_{s_1\bm{\mu}}^{s_2\bm{\nu};s_3\bm{\rho}}(p_1)=&~P^{(s_1)}_{\bm{\mu};\bm{\alpha}\bm{\gamma}}(p_1)\,g_{\bm{\beta},\bm{\beta}'}\notag\\
			&\times P^{(s_2)\bm{\nu};\bm{\alpha}\bm{\beta}}(p_1)\,P^{(s_3)\bm{\gamma}\bm{\beta}';\bm{\rho}}(p_1).\label{eq:AMCT-2B}
		\end{align}
		The half-integer-spin projection tensors involved in Eqs.~\eqref{eq:AMCT-2A} and \eqref{eq:AMCT-2B} can be constructed from the integer-spin projection tensors and the Dirac $\gamma$-matrices; for $n\in\mathbb{N}$ they read \cite{Behrends:1957rup,Huang:2003jg}
		\begin{equation}
			P_{\bm{\mu};\bm{\nu}}^{\left(n+\frac{1}{2}\right)}(p_1)=\frac{n+1}{2 n+3} \gamma^\alpha \gamma^\beta P_{\bm{\mu}\alpha;\bm{\nu}\beta}^{(n+1)}(p_1).
		\end{equation}
		The numbers of indices in the index sets $\bm{\mu},\bm{\nu},\bm{\rho}$ and $\bm{\alpha},\bm{\beta}^{(\prime)},\bm{\gamma}$ in Eqs.~\eqref{eq:AMCT-2A} and \eqref{eq:AMCT-2B} are $s_1,s_2-\frac12,s_3-\frac12$ and $a_\mathrm{II},b_\mathrm{II},c_\mathrm{II}$, respectively. 
		The triplet $(a_\mathrm{II},b_\mathrm{II},c_\mathrm{II})$ characterizing the numbers of indices reads
		\begin{align}
			a_\mathrm{II}=&~\frac{s_1+s_2-s_3+\delta-1}{2},
			\label{eq:aII}\\
			b_\mathrm{II}=&~\frac{s_2+s_3-s_1-\delta}{2},
			\label{eq:bII}\\
			c_\mathrm{II}=&~\frac{s_1+s_3-s_2+\delta-1}{2}.
			\label{eq:cII}
		\end{align}

		\item	Case III
		
		$s_1,s_3$ are both half-integer spins (in this case $s_2$ is necessarily an integer spin).
		For $\delta=0$ the coupling structure tensor reads
		\begin{align}
			P_{s_1\bm{\mu}}^{s_2\bm{\nu};s_3\bm{\rho}}(p_1)=&~P^{(s_1)}_{\bm{\mu};\bm{\alpha}\bm{\gamma}}(p_1)\,\gamma_5\gamma_{\lambda}\,P^{(s_3)\bm{\gamma}\bm{\beta}';\bm{\rho}}(p_1)\notag\\&\times P^{(s_2)\bm{\nu};\bm{\alpha}\bm{\beta}\lambda}(p_1)\,g_{\bm{\beta},\bm{\beta}'}.\label{eq:AMCT-3A}
		\end{align}
		For $\delta=1$ it reads
		\begin{align}
			P_{s_1\bm{\mu}}^{s_2\bm{\nu};s_3\bm{\rho}}(p_1)=&~P^{(s_1)}_{\bm{\mu};\bm{\alpha}\bm{\gamma}}(p_1)\,P^{(s_3)\bm{\gamma}\bm{\beta}';\bm{\rho}}(p_1)\notag\\
			&\times P^{(s_2)\bm{\nu};\bm{\alpha}\bm{\beta}}(p_1)\,g_{\bm{\beta},\bm{\beta}'}.\label{eq:AMCT-3B}
		\end{align}
		The numbers of indices in the index sets $\bm{\mu},\bm{\nu},\bm{\rho}$ and $\bm{\alpha},\bm{\beta}^{(\prime)},\bm{\gamma}$ in Eqs.~\eqref{eq:AMCT-3A} and \eqref{eq:AMCT-3B} are $s_1-\frac12,s_2,s_3-\frac12$ and $a_\mathrm{III},b_\mathrm{III},c_\mathrm{III}$, respectively. 
		The triplet $(a_\mathrm{III},b_\mathrm{III},c_\mathrm{III})$ characterizing the numbers of indices reads
		\begin{align}
			a_\mathrm{III}=&~\frac{s_1+s_2-s_3+\delta-1}{2},
			\label{eq:aIII}\\
			b_\mathrm{III}=&~\frac{s_2+s_3-s_1+\delta-1}{2},
			\label{eq:bIII}\\
			c_\mathrm{III}=&~\frac{s_1+s_3-s_2-\delta}{2}.
			\label{eq:cIII}
		\end{align}
	\end{itemize}
	
	It should be pointed out that the angular momentum coupling tensors given in Eqs.~\eqref{eq:AMCT-2A} and \eqref{eq:AMCT-2B} apply only to the coupling between a fermionic row spinor $\bar{u}_{\bm{\nu}}(p_2)$ and a fermionic column spinor $u_{\bm{\rho}}(p_3)$. 
	When other types of spinor couplings are involved, the spinors must be replaced accordingly; similarly, the angular momentum coupling tensors given in Eqs.~\eqref{eq:AMCT-3A} and \eqref{eq:AMCT-3B} apply only to the coupling between a fermionic row spinor $\bar{u}^{\bm{\mu}}(p_1)$ and a fermionic column spinor $u_{\bm{\rho}}(p_3)$, and the spinors must also be replaced accordingly when other types of spinor couplings are involved. 
	The specific replacement rules, which follow from the standard charge-conjugation relations of the Dirac spinors \cite{Weinberg:1995mt}, are given in Table~\ref{tab:Dirac-spinor-replace-rule}.

	\begin{table*}
		\centering
		\caption{Replacement rules for the spinor wave functions in the angular momentum coupling tensors $P_{s_1\bm{\mu}}^{s_2\bm{\nu};s_3\bm{\rho}}(p_1)$ involving fermion pairs, where $C=i\gamma^0\gamma^2$ is the charge conjugation matrix in the Dirac spinor representation. The replacement rules also apply to the couplings between the RSWFs introduced above.}
		\begin{tabular}{ccccc}
			\hline\hline
			& ~~~fermion creation~~~ & ~~~fermion annihilation~~~ & ~~~antifermion creation~~~ & ~~~antifermion annihilation~~~ \\
			\hline
			row spinor-1 & $\bar{u}^{\bm{\mu}}(p_1)$ & $\left[u^{\bm{\mu}}(p_1)\right]^{\mathrm{T}}\gamma_5C$ & $\left[v^{\bm{\mu}}(p_1)\right]^{\mathrm{T}}C$ & $\bar{v}^{\bm{\mu}}(p_1)\gamma_5$  \\
			row spinor-2 & $\bar{u}_{\bm{\nu}}(p_2)$ & $\left[u_{\bm{\nu}}(p_2)\right]^{\mathrm{T}}\gamma_5C$ & $\left[v_{\bm{\nu}}(p_2)\right]^{\mathrm{T}}C$ & $\bar{v}_{\bm{\nu}}(p_2)\gamma_5$  \\
			column spinor-3 & $\gamma_5C\left[\bar{u}_{\bm{\rho}}(p_3)\right]^{\mathrm{T}}$ & $u_{\bm{\rho}}(p_3)$ & $\gamma_5v_{\bm{\rho}}(p_3)$ & $C\left[\bar{v}_{\bm{\rho}}(p_3)\right]^{\mathrm{T}}$ \\
			\hline\hline
		\end{tabular}
		\label{tab:Dirac-spinor-replace-rule}
	\end{table*}

	The three cases I, II, and III presented in this section correspond, respectively, to the situations in which the number of half-integer spins among the three external particles is 0, 2 (the final-state pair), and 2 (the initial particle and particle 3); the remaining combinations either require an interchange of the labels of particles 2 and 3, or violate the condition of angular momentum conservation.

	\runin{\bf Covariant on-shell PWAs}
	Combining the RSWFs introduced in the preceding section with the coupling structure tensors constructed above yields the manifestly covariant form of the $\mathcal{H}$-type amplitude of any two-body decay process.
	In this section, we present the explicit construction of the $\mathcal{H}$-type amplitudes of two-body decay processes in the RS form. 
	For later convenience, we introduce $\delta_1=s_1+L+S\mod{2}$ and $\delta_2=S+s_2+s_3\mod{2}$. 
	Note that, constrained by angular momentum conservation, the ranges of the quantum numbers $S$ and $L$ are
	\begin{align}
		S\in&\{|s_2-s_3|,|s_2-s_3|+1,\cdots,s_2+s_3\},\\
		L\in&\{|s_1-S|,|s_1-S|+1,\cdots,s_1+S\}.
	\end{align}

	We first consider two-body decay processes involving only bosons. 
	Substituting Eq.~\eqref{eq:Gamma-PPq} into Eq.~\eqref{eq:pwamp-H-abf}, the manifestly covariant form of the $\mathcal{H}$-type amplitude in this case reads
	\begin{align}
		\mathcal{H}_{L,S}^{(\mathrm{I})}=&~\varepsilon^{\bm{\mu}}(p_1)\,P_{s_1\bm{\mu}}^{L\bm{\xi};S\bm{\sigma}}(p_1)\,q_{\bm{\xi}}\,P_{S\bm{\sigma}}^{s_2\bm{\nu};s_3\bm{\rho}}(p_1)\notag\\
		&\times\varepsilon^*_{\bm{\nu}}(p_2|p_1)\,\varepsilon^*_{\bm{\rho}}(p_3|p_1).
	\end{align}
	Depending on the values of $\delta_1$ and $\delta_2$, the amplitude $\mathcal{H}_{L,S}^{(\mathrm{I})}$ can be further divided into four types, denoted as $\mathcal{H}_{L,S}^{(\mathrm{I}\text{-}\delta_1\delta_2)}$. With Eqs.~\eqref{eq:AMCT-1A} and \eqref{eq:AMCT-1B}, the manifestly covariant forms of $\mathcal{H}_{L,S}^{(\mathrm{I}\text{-}\delta_1\delta_2)}$ are obtained.

	Next we consider two-body decay processes with a fermion-antifermion pair in the final state. 
	Substituting Eq.~\eqref{eq:Gamma-PPq} into Eq.~\eqref{eq:pwamp-H-abf}, the manifestly covariant form of the $\mathcal{H}$-type amplitude in this case reads
	\begin{align}
		\mathcal{H}_{L,S}^{(\mathrm{II})}=&~\varepsilon^{\bm{\mu}}(p_1)\,P_{s_1\bm{\mu}}^{L\bm{\xi};S\bm{\sigma}}(p_1)\,q_{\bm{\xi}}
		\notag\\&\times
		\bar{u}_{\bm{\nu}}(p_2|p_1)\,P_{S\bm{\sigma}}^{s_2\bm{\nu};s_3\bm{\rho}}(p_1)\,\gamma_5v_{\bm{\rho}}(p_3|p_1),
	\end{align}
	where, according to the replacement rules in Table~\ref{tab:Dirac-spinor-replace-rule}, we have replaced the relative spin wave function $u_{\bm{\rho}}(p_3|p_1)$ of particle 3 by $\gamma_5v_{\bm{\rho}}(p_3|p_1)$, since we are now describing the production of a fermion-antifermion pair. 
	Depending on the values of $\delta_1$ and $\delta_2$, the amplitude $\mathcal{H}_{L,S}^{(\mathrm{II})}$ can also be divided into four types, denoted $\mathcal{H}_{L,S}^{(\mathrm{II}\text{-}\delta_1\delta_2)}$; with Eqs.~\eqref{eq:AMCT-1A}, \eqref{eq:AMCT-1B}, \eqref{eq:AMCT-2A}, and \eqref{eq:AMCT-2B}, their manifestly covariant forms are obtained.

	Finally we consider two-body decay processes with a fermion in the initial state; we take particle 1 and particle 3 to be fermions. 
	Substituting Eq.~\eqref{eq:Gamma-PPq} into Eq.~\eqref{eq:pwamp-H-abf}, the manifestly covariant form of the $\mathcal{H}$-type amplitude in this case reads
	\begin{align}
		\mathcal{H}_{L,S}^{(\mathrm{III})}=&~\varepsilon^{*}_{\bm{\nu}}(p_2|p_1)\,q_{\bm{\xi}}\,\left[u^{\bm{\mu}}(p_1)\right]^{\mathrm{T}}\gamma_5 C\,P_{s_1\bm{\mu}}^{L\bm{\xi};S\bm{\sigma}}(p_1)
		\notag\\&\times 
		P_{S\bm{\sigma}}^{s_2\bm{\nu};s_3\bm{\rho}}(p_1)\,\gamma_5C\left[\bar{u}_{\bm{\rho}}(p_3|p_1)\right]^{\mathrm{T}},
	\end{align}
	where, according to the replacement rules in Table~\ref{tab:Dirac-spinor-replace-rule}, we have replaced the RSWFs $\bar{u}^{\bm{\mu}}(p_1)$ and $u_{\bm{\rho}}(p_3|p_1)$ of particles 1 and 3 by $\left[u^{\bm{\mu}}(p_1)\right]^{\mathrm{T}}\gamma_5 C$ and $\gamma_5C\left[\bar{u}_{\bm{\rho}}(p_3|p_1)\right]^{\mathrm{T}}$, respectively, since we are now describing a process in which an initial fermion annihilates and a final fermion is created. 
	Depending on the values of $\delta_1$ and $\delta_2$, the amplitude $\mathcal{H}_{L,S}^{(\mathrm{III})}$ can also be divided into four types, denoted $\mathcal{H}_{L,S}^{(\mathrm{III}\text{-}\delta_1\delta_2)}$; with Eqs.~\eqref{eq:AMCT-3A} and \eqref{eq:AMCT-3B}, their manifestly covariant forms are obtained.

	\begin{table*}
		\centering
		\small
		\caption{
			Explicit expressions of the $\mathcal{H}$-type partial-wave amplitudes $\mathcal{H}_{L,S}^{(\mathrm{X}\text{-}\delta_1\delta_2)}$ ($\mathrm{X}=\mathrm{I},\mathrm{II},\mathrm{III}$) of two-body decay processes: case I (only bosons), case II (a fermion-antifermion pair in the final state), and case III (a fermion in the initial state, with particles 1 and 3 taken as fermions). 
			The symbol $\tilde{t}^{(L)}(p_1,q)$ denotes the orbital angular momentum tensor \cite{Zemach:1965ycj}; for its covariant form and the explicit expressions for the lowest values of $L$, see, e.g., Refs.~\cite{Zou:2002ar,Dulat:2005in,Dulat:2011rn,Huang:2026egv}. 
			For example, $\tilde{t}^{(L)\bm{\alpha}\bm{\beta}}(p_1,q)=P^{(L)\bm{\alpha}\bm{\beta};\bm{\xi}}(p_1)\,q_{\bm{\xi}}$; in deriving the case-III formulas we have used the relation $\gamma_5C\gamma_\mu\gamma_5C=-\gamma_\mu^{\mathrm{T}}$. 
			The numbers of indices in the index sets $\bm{\alpha},\bm{\beta},\bm{\gamma}$ and $\bm{\alpha}',\bm{\beta}',\bm{\gamma}'$ are determined, based on the angular momentum triplets $(s_1,L,S)$ and $(S,s_2,s_3)$ respectively, as follows: in case I both blocks follow Eqs.~\eqref{eq:aI}-\eqref{eq:cI}; in case II the first block follows Eqs.~\eqref{eq:aI}-\eqref{eq:cI} and the second block follows Eqs.~\eqref{eq:aII}-\eqref{eq:cII}; in case III both blocks follow Eqs.~\eqref{eq:aIII}-\eqref{eq:cIII}.
		}
		\begin{tabular}{rl}
			\hline\hline
			\multicolumn{2}{l}{Case I: two-body decay processes involving only bosons}\\
			\hline
			$\mathcal{H}_{L,S}^{(\mathrm{I}\text{-}00)}=$ & $\varepsilon_{\bm{\alpha}}^{\bm{\gamma}}(p_1)\,\tilde{t}^{(L)\bm{\alpha}\bm{\beta}}(p_1,q)\,P^{(S)}_{\bm{\beta}\bm{\gamma};\bm{\alpha}'\bm{\gamma}'}(p_1)\,\varepsilon^{*~\bm{\alpha}'}_{\bm{\beta}'}(p_2|p_1)\,\varepsilon^{*\bm{\beta}'\bm{\gamma}'}(p_3|p_1)$\\
			$\mathcal{H}_{L,S}^{(\mathrm{I}\text{-}01)}=$ & $i\,\varepsilon_{\bm{\alpha}}^{\bm{\gamma}}(p_1)\,\tilde{t}^{(L)\bm{\alpha}\bm{\beta}}(p_1,q)\,P^{(S)}_{\bm{\beta}\bm{\gamma};\bm{\alpha}'\bm{\gamma}'\lambda}(p_1)\,\tilde{\epsilon}^{\lambda}_{\kappa\tau}(p_1)\,\varepsilon^{*~\bm{\alpha}'\kappa}_{\bm{\beta}'}(p_2|p_1)\,\varepsilon^{*\bm{\beta}'\bm{\gamma}'\tau}(p_3|p_1)$\\
			$\mathcal{H}_{L,S}^{(\mathrm{I}\text{-}10)}=$ & $i\,\varepsilon_{\bm{\alpha}}^{\bm{\gamma}\mu}(p_1)\,\tilde{t}^{(L)\bm{\alpha}\bm{\beta}}_{\nu}(p_1,q)\,P^{(S)}_{\bm{\beta}\bm{\gamma}\rho;\bm{\alpha}'\bm{\gamma}'}(p_1)\,\tilde{\epsilon}_{\mu}^{\nu\rho}(p_1)\,\varepsilon^{*~\bm{\alpha}'}_{\bm{\beta}'}(p_2|p_1)\,\varepsilon^{*\bm{\beta}'\bm{\gamma}'}(p_3|p_1)$\\
			$\mathcal{H}_{L,S}^{(\mathrm{I}\text{-}11)}=$ & $-\varepsilon_{\bm{\alpha}}^{\bm{\gamma}\mu}(p_1)\,\tilde{t}^{(L)\bm{\alpha}\bm{\beta}}_{\nu}(p_1,q)\,P^{(S)}_{\bm{\beta}\bm{\gamma}\rho;\bm{\alpha}'\bm{\gamma}'\lambda}(p_1)\,\tilde{\epsilon}_{\mu}^{\nu\rho}(p_1)\,\tilde{\epsilon}^{\lambda}_{\kappa\tau}(p_1)\,\varepsilon^{*~\bm{\alpha}'\kappa}_{\bm{\beta}'}(p_2|p_1)\,\varepsilon^{*\bm{\beta}'\bm{\gamma}'\tau}(p_3|p_1)$\\
			\hline
			\multicolumn{2}{l}{Case II: two-body decay processes with a fermion-antifermion pair in the final state}\\
			\hline
			$\mathcal{H}_{L,S}^{(\mathrm{II}\text{-}00)}=$ & $-\varepsilon_{\bm{\alpha}}^{\bm{\gamma}}(p_1)\,\tilde{t}^{(L)\bm{\alpha}\bm{\beta}}(p_1,q)\,P^{(S)}_{\bm{\beta}\bm{\gamma};\bm{\alpha}'\bm{\gamma}'\lambda}(p_1)\,\bar{u}^{\bm{\alpha}'\bm{\beta}'}(p_2|p_1)\,\gamma^{\lambda}\,v^{\bm{\gamma}'}_{\bm{\beta}'}(p_3|p_1)$\\
			$\mathcal{H}_{L,S}^{(\mathrm{II}\text{-}01)}=$ & $\varepsilon_{\bm{\alpha}}^{\bm{\gamma}}(p_1)\,\tilde{t}^{(L)\bm{\alpha}\bm{\beta}}(p_1,q)\,P^{(S)}_{\bm{\beta}\bm{\gamma};\bm{\alpha}'\bm{\gamma}'}(p_1)\,\bar{u}^{\bm{\alpha}'\bm{\beta}'}(p_2|p_1)\,\gamma_5\,v^{\bm{\gamma}'}_{\bm{\beta}'}(p_3|p_1)$\\
			$\mathcal{H}_{L,S}^{(\mathrm{II}\text{-}10)}=$ & $-i\,\varepsilon_{\bm{\alpha}}^{\bm{\gamma}\mu}(p_1)\,\tilde{t}^{(L)\bm{\alpha}\bm{\beta}}_{\nu}(p_1,q)\,P^{(S)}_{\bm{\beta}\bm{\gamma}\rho;\bm{\alpha}'\bm{\gamma}'\lambda}(p_1)\,\tilde{\epsilon}_{\mu}^{\nu\rho}(p_1)\,\bar{u}^{\bm{\alpha}'\bm{\beta}'}(p_2|p_1)\,\gamma^{\lambda}\,v^{\bm{\gamma}'}_{\bm{\beta}'}(p_3|p_1)$\\
			$\mathcal{H}_{L,S}^{(\mathrm{II}\text{-}11)}=$ & $i\varepsilon_{\bm{\alpha}}^{\bm{\gamma}\mu}(p_1)\,\tilde{t}^{(L)\bm{\alpha}\bm{\beta}}_{\nu}(p_1,q)\,P^{(S)}_{\bm{\beta}\bm{\gamma}\rho;\bm{\alpha}'\bm{\gamma}'}(p_1)\,\tilde{\epsilon}_{\mu}^{\nu\rho}(p_1)\,\bar{u}^{\bm{\alpha}'\bm{\beta}'}(p_2|p_1)\,\gamma_5\,v^{\bm{\gamma}'}_{\bm{\beta}'}(p_3|p_1)$\\
			\hline
			\multicolumn{2}{l}{Case III: two-body decay processes with a fermion in the initial state}\\
			\hline
			$\mathcal{H}_{L,S}^{(\mathrm{III}\text{-}00)}=$ & $\varepsilon_{\bm{\beta}'\mu}^{*\bm{\alpha}'}(p_2|p_1)\,\tilde{t}^{(L)\bm{\alpha}\bm{\beta}}_{\nu}(p_1,q)\,\bar{u}^{\bm{\beta}'\bm{\gamma}'}(p_3|p_1)\,\gamma^{\mu}\,P^{(S)}_{\bm{\alpha}'\bm{\gamma}';\bm{\beta}\bm{\gamma}}(p_1)\,\gamma^\nu\,u^{\bm{\gamma}}_{\bm{\alpha}}(p_1)$\\
			$\mathcal{H}_{L,S}^{(\mathrm{III}\text{-}01)}=$ & $\varepsilon_{\bm{\beta}'}^{*\bm{\alpha}'}(p_2|p_1)\,\tilde{t}^{(L)\bm{\alpha}\bm{\beta}}_{\mu}(p_1,q)\,\bar{u}^{\bm{\beta}'\bm{\gamma}'}(p_3|p_1)\,P^{(S)}_{\bm{\alpha}'\bm{\gamma}';\bm{\beta}\bm{\gamma}}(p_1)\,\gamma_5\gamma^\mu\,u^{\bm{\gamma}}_{\bm{\alpha}}(p_1)$\\
			$\mathcal{H}_{L,S}^{(\mathrm{III}\text{-}10)}=$ & $\varepsilon_{\bm{\beta}'\mu}^{*\bm{\alpha}'}(p_2|p_1)\,\tilde{t}^{(L)\bm{\alpha}\bm{\beta}}(p_1,q)\,\bar{u}^{\bm{\beta}'\bm{\gamma}'}(p_3|p_1)\,\gamma_5\gamma^\mu\,P^{(S)}_{\bm{\alpha}'\bm{\gamma}';\bm{\beta}\bm{\gamma}}(p_1)\,u^{\bm{\gamma}}_{\bm{\alpha}}(p_1)$\\
			$\mathcal{H}_{L,S}^{(\mathrm{III}\text{-}11)}=$ & $-\varepsilon_{\bm{\beta}'}^{*\bm{\alpha}'}(p_2|p_1)\,\tilde{t}^{(L)\bm{\alpha}\bm{\beta}}(p_1,q)\,\bar{u}^{\bm{\beta}'\bm{\gamma}'}(p_3|p_1)\,P^{(S)}_{\bm{\alpha}'\bm{\gamma}';\bm{\beta}\bm{\gamma}}(p_1)\,u^{\bm{\gamma}}_{\bm{\alpha}}(p_1)$\\
			\hline\hline
		\end{tabular} \label{tab:pwamp-H-exp}
	\end{table*}

	The explicit expressions of the four types of amplitudes in the three cases are collected in Table~\ref{tab:pwamp-H-exp}.
	For reference, for two-body decay processes with all external spins no larger than 2, the Supplemental Material accompanying this Letter lists the numbers of indices in the index sets $\bm{\alpha},\bm{\beta},\bm{\gamma}$ and $\bm{\alpha}',\bm{\beta}',\bm{\gamma}'$ for all angular-momentum-allowed $L$-$S$ combinations. 
	Combining the formulas in Table~\ref{tab:pwamp-H-exp} with these numbers of indices, the explicit expressions of the $\mathcal{H}$-type partial-wave amplitudes of the corresponding processes in the RS form can be written down very conveniently. 
	The Supplemental Material also contains explicit examples of the construction of the covariant on-shell partial-wave amplitudes, together with comparisons with Refs.~\cite{Zou:2002ar,Zou:2002yy}.
	
	\runin{\bf Summary and outlook}
	In this Letter, we have constructed manifestly covariant
	$\mathcal{H}$-type partial-wave amplitudes for two-body decays of massive particles with arbitrary spins in the RS formalism.
	By introducing RSWFs, our formulation separates covariant spin transport from orbital-spin coupling, placing the pure-spin ($\mathcal{H}$-type) and general-spin ($\mathcal{C}$-type) schemes on an equal covariant footing.
	The representation-independent definition in Eq.~\eqref{eq:def-relative-swf} also provides a starting point for relating the RS construction to the covariant canonical-spinor scheme~\cite{Huang:2026egv} and the spinor-helicity representation~\cite{Arkani-Hamed:2017jhn,Elvang:2013cua}.

	For phenomenological applications, these amplitudes provide building blocks for cascade-decay analyses in the isobar 	model~\cite{Herndon1975Isobar, Krinner:2017}.
	Their explicit tensor form is suitable for implementation
	in automated PWA frameworks, such as FDCPWA~\cite{Wang:2004fdc}, TFPWA~\cite{tfpwa}, CTPWA~\cite{Xiang:2026nwt} and so on.
	
	Beyond tree-level amplitude analyses, our manifestly covariant formulation can be extended to off-shell amplitudes, opening further avenues in hadron dynamics and spectroscopy.
	The off-shell extension offers a route to loop calculations
	in effective field theories.
	In analyses based on dispersion relations, the on-shell
	amplitudes provide a covariant basis for organizing two-body
	intermediate states with definite $L$-$S$ quantum numbers
	on unitarity cuts.
	At the quark-gluon level, these covariant coupling tensors may also guide the construction of interpolating operators organized by ${}^{2S+1}L_J$ for QCD sum rules~\cite{Dai1997ExcitedMesons, Chen2015PwaveBaryons}
	and, after subduction onto lattice irreducible representations, lattice QCD~\cite{Dudek2010MesonOperators, Edwards2011BaryonOperators}.

	\section*{Acknowledgments}
	This work is supported by the National Natural Science Foundation of China under Grants No.~12405100, 12547111 and 12221005, 
	and by the Chinese Academy of Sciences under Grant No. YSBR-101,
	and by the National Key Research and Development Program of
	China under Contract No. 2025YFA1613900.
	
	\bibliography{refs.bib}

	\clearpage
	\onecolumngrid
	\section*{Supplemental Material}
	\label{sec:supplemental}
	
	\subsection*{A. Numbers of elements of the index sets}
	
	The table below lists the numbers of elements of the index sets $\bm{\alpha},\bm{\beta},\bm{\gamma}$ and $\bm{\alpha}',\bm{\beta}',\bm{\gamma}'$ for all angular-momentum-allowed $L$-$S$ combinations of two-body decay processes $1\to2+3$ with all external spins no larger than $2$.
	
	\begin{footnotesize}
	\setlength{\tabcolsep}{4pt} 
	\setlength{\LTcapwidth}{\textwidth} 
	\renewcommand{\arraystretch}{1.45} 
	\makeatletter\set@arstrutbox\makeatother 
	\begin{longtable}{ccccc@{\hspace{1.2em}}|@{\hspace{1.2em}}ccccc}
		\caption{Numbers of elements of the index sets for all angular-momentum-allowed $L$-$S$ combinations of two-body decay processes $1\to2+3$ with all external spins no larger than $2$. The second column lists $(L,S)$; in the type column $\mathrm{X}\text{-}\delta_1\delta_2$, $\delta_1\equiv s_1+L+S\bmod 2$ and $\delta_2\equiv S+s_2+s_3\bmod 2$ correspond one-to-one to the formulas with the same labels in Table~II of the main text. The columns $\#(\bm{\alpha},\bm{\beta},\bm{\gamma})$ and $\#(\bm{\alpha}^{\prime},\bm{\beta}^{\prime},\bm{\gamma}^{\prime})$ give, in the form $(a,b,c)$ and $(a^{\prime},b^{\prime},c^{\prime})$, the numbers of elements of the index sets $\bm{\alpha},\bm{\beta},\bm{\gamma}$ and $\bm{\alpha}^{\prime},\bm{\beta}^{\prime},\bm{\gamma}^{\prime}$ based on the angular momentum triplets $(s_1,L,S)$ and $(S,s_2,s_3)$, respectively; the symbol ``$-$" denotes the triple of index counts $(0,0,0)$, meaning that all three index sets in the corresponding column are empty. For each case the rows are given in two side-by-side groups, the left group coming first. The numbers of elements are computed according to the following rules: in case I both index blocks follow Eqs.~\eqref{eq:aI}--\eqref{eq:cI}; in case II the first index block follows Eqs.~\eqref{eq:aI}--\eqref{eq:cI} and the second one follows Eqs.~\eqref{eq:aII}--\eqref{eq:cII}; in case III both index blocks follow Eqs.~\eqref{eq:aIII}--\eqref{eq:cIII}. With this table, the explicit expressions of the corresponding $\mathcal{H}$-type partial-wave amplitudes can be written down directly. For processes in which the final fermion is particle 2 ($s_2$ half-integer and $s_3$ integer), interchange the labels of particles 2 and 3 and consult this table as case III.}\label{tab:index-counts}\\
		\hline\hline
		$(s_1,s_2,s_3)$ & $(L,S)$ & Type & {\scriptsize $\#(\bm{\alpha},\bm{\beta},\bm{\gamma})$} & {\scriptsize $\#(\bm{\alpha}^{\prime},\bm{\beta}^{\prime},\bm{\gamma}^{\prime})$} & $(s_1,s_2,s_3)$ & $(L,S)$ & Type & {\scriptsize $\#(\bm{\alpha},\bm{\beta},\bm{\gamma})$} & {\scriptsize $\#(\bm{\alpha}^{\prime},\bm{\beta}^{\prime},\bm{\gamma}^{\prime})$} \\
		\hline
		\endfirsthead
		\hline
		$(s_1,s_2,s_3)$ & $(L,S)$ & Type & {\scriptsize $\#(\bm{\alpha},\bm{\beta},\bm{\gamma})$} & {\scriptsize $\#(\bm{\alpha}^{\prime},\bm{\beta}^{\prime},\bm{\gamma}^{\prime})$} & $(s_1,s_2,s_3)$ & $(L,S)$ & Type & {\scriptsize $\#(\bm{\alpha},\bm{\beta},\bm{\gamma})$} & {\scriptsize $\#(\bm{\alpha}^{\prime},\bm{\beta}^{\prime},\bm{\gamma}^{\prime})$} \\
		\hline
		\endhead
		\hline
		\multicolumn{10}{r}{\footnotesize continued on next page} \\
		\endfoot
		\hline\hline
		\endlastfoot
		\hline
		\multicolumn{10}{c}{\textbf{Case I: $s_1,s_2,s_3$ all integer spins}} \\
		\hline
		$(0,\,0,\,0)$ & $(0,\,0)$ & $\mathrm{I}\text{-}00$ & $-$ & $-$ & $(2,\,0,\,0)$ & $(2,\,0)$ & $\mathrm{I}\text{-}00$ & $(2,\,0,\,0)$ & $-$ \\
		$(0,\,0,\,1)$ & $(1,\,1)$ & $\mathrm{I}\text{-}00$ & $(0,\,1,\,0)$ & $(0,\,0,\,1)$ & $(2,\,0,\,1)$ & $(1,\,1)$ & $\mathrm{I}\text{-}00$ & $(1,\,0,\,1)$ & $(0,\,0,\,1)$ \\
		$(0,\,0,\,2)$ & $(2,\,2)$ & $\mathrm{I}\text{-}00$ & $(0,\,2,\,0)$ & $(0,\,0,\,2)$ &  & $(2,\,1)$ & $\mathrm{I}\text{-}10$ & $(1,\,0,\,0)$ & $(0,\,0,\,1)$ \\
		$(0,\,1,\,0)$ & $(1,\,1)$ & $\mathrm{I}\text{-}00$ & $(0,\,1,\,0)$ & $(1,\,0,\,0)$ &  & $(3,\,1)$ & $\mathrm{I}\text{-}00$ & $(2,\,1,\,0)$ & $(0,\,0,\,1)$ \\
		$(0,\,1,\,1)$ & $(0,\,0)$ & $\mathrm{I}\text{-}00$ & $-$ & $(0,\,1,\,0)$ & $(2,\,0,\,2)$ & $(0,\,2)$ & $\mathrm{I}\text{-}00$ & $(0,\,0,\,2)$ & $(0,\,0,\,2)$ \\
		& $(1,\,1)$ & $\mathrm{I}\text{-}01$ & $(0,\,1,\,0)$ & $-$ &  & $(1,\,2)$ & $\mathrm{I}\text{-}10$ & $(0,\,0,\,1)$ & $(0,\,0,\,2)$ \\
		& $(2,\,2)$ & $\mathrm{I}\text{-}00$ & $(0,\,2,\,0)$ & $(1,\,0,\,1)$ &  & $(2,\,2)$ & $\mathrm{I}\text{-}00$ & $(1,\,1,\,1)$ & $(0,\,0,\,2)$ \\
		$(0,\,1,\,2)$ & $(1,\,1)$ & $\mathrm{I}\text{-}00$ & $(0,\,1,\,0)$ & $(0,\,1,\,1)$ &  & $(3,\,2)$ & $\mathrm{I}\text{-}10$ & $(1,\,1,\,0)$ & $(0,\,0,\,2)$ \\
		& $(2,\,2)$ & $\mathrm{I}\text{-}01$ & $(0,\,2,\,0)$ & $(0,\,0,\,1)$ &  & $(4,\,2)$ & $\mathrm{I}\text{-}00$ & $(2,\,2,\,0)$ & $(0,\,0,\,2)$ \\
		& $(3,\,3)$ & $\mathrm{I}\text{-}00$ & $(0,\,3,\,0)$ & $(1,\,0,\,2)$ & $(2,\,1,\,0)$ & $(1,\,1)$ & $\mathrm{I}\text{-}00$ & $(1,\,0,\,1)$ & $(1,\,0,\,0)$ \\
		$(0,\,2,\,0)$ & $(2,\,2)$ & $\mathrm{I}\text{-}00$ & $(0,\,2,\,0)$ & $(2,\,0,\,0)$ &  & $(2,\,1)$ & $\mathrm{I}\text{-}10$ & $(1,\,0,\,0)$ & $(1,\,0,\,0)$ \\
		$(0,\,2,\,1)$ & $(1,\,1)$ & $\mathrm{I}\text{-}00$ & $(0,\,1,\,0)$ & $(1,\,1,\,0)$ &  & $(3,\,1)$ & $\mathrm{I}\text{-}00$ & $(2,\,1,\,0)$ & $(1,\,0,\,0)$ \\
		& $(2,\,2)$ & $\mathrm{I}\text{-}01$ & $(0,\,2,\,0)$ & $(1,\,0,\,0)$ & $(2,\,1,\,1)$ & $(2,\,0)$ & $\mathrm{I}\text{-}00$ & $(2,\,0,\,0)$ & $(0,\,1,\,0)$ \\
		& $(3,\,3)$ & $\mathrm{I}\text{-}00$ & $(0,\,3,\,0)$ & $(2,\,0,\,1)$ &  & $(1,\,1)$ & $\mathrm{I}\text{-}01$ & $(1,\,0,\,1)$ & $-$ \\
		$(0,\,2,\,2)$ & $(0,\,0)$ & $\mathrm{I}\text{-}00$ & $-$ & $(0,\,2,\,0)$ &  & $(2,\,1)$ & $\mathrm{I}\text{-}11$ & $(1,\,0,\,0)$ & $-$ \\
		& $(1,\,1)$ & $\mathrm{I}\text{-}01$ & $(0,\,1,\,0)$ & $(0,\,1,\,0)$ &  & $(3,\,1)$ & $\mathrm{I}\text{-}01$ & $(2,\,1,\,0)$ & $-$ \\
		& $(2,\,2)$ & $\mathrm{I}\text{-}00$ & $(0,\,2,\,0)$ & $(1,\,1,\,1)$ &  & $(0,\,2)$ & $\mathrm{I}\text{-}00$ & $(0,\,0,\,2)$ & $(1,\,0,\,1)$ \\
		& $(3,\,3)$ & $\mathrm{I}\text{-}01$ & $(0,\,3,\,0)$ & $(1,\,0,\,1)$ &  & $(1,\,2)$ & $\mathrm{I}\text{-}10$ & $(0,\,0,\,1)$ & $(1,\,0,\,1)$ \\
		& $(4,\,4)$ & $\mathrm{I}\text{-}00$ & $(0,\,4,\,0)$ & $(2,\,0,\,2)$ &  & $(2,\,2)$ & $\mathrm{I}\text{-}00$ & $(1,\,1,\,1)$ & $(1,\,0,\,1)$ \\
		$(1,\,0,\,0)$ & $(1,\,0)$ & $\mathrm{I}\text{-}00$ & $(1,\,0,\,0)$ & $-$ &  & $(3,\,2)$ & $\mathrm{I}\text{-}10$ & $(1,\,1,\,0)$ & $(1,\,0,\,1)$ \\
		$(1,\,0,\,1)$ & $(0,\,1)$ & $\mathrm{I}\text{-}00$ & $(0,\,0,\,1)$ & $(0,\,0,\,1)$ &  & $(4,\,2)$ & $\mathrm{I}\text{-}00$ & $(2,\,2,\,0)$ & $(1,\,0,\,1)$ \\
		& $(1,\,1)$ & $\mathrm{I}\text{-}10$ & $-$ & $(0,\,0,\,1)$ & $(2,\,1,\,2)$ & $(1,\,1)$ & $\mathrm{I}\text{-}00$ & $(1,\,0,\,1)$ & $(0,\,1,\,1)$ \\
		& $(2,\,1)$ & $\mathrm{I}\text{-}00$ & $(1,\,1,\,0)$ & $(0,\,0,\,1)$ &  & $(2,\,1)$ & $\mathrm{I}\text{-}10$ & $(1,\,0,\,0)$ & $(0,\,1,\,1)$ \\
		$(1,\,0,\,2)$ & $(1,\,2)$ & $\mathrm{I}\text{-}00$ & $(0,\,1,\,1)$ & $(0,\,0,\,2)$ &  & $(3,\,1)$ & $\mathrm{I}\text{-}00$ & $(2,\,1,\,0)$ & $(0,\,1,\,1)$ \\
		& $(2,\,2)$ & $\mathrm{I}\text{-}10$ & $(0,\,1,\,0)$ & $(0,\,0,\,2)$ &  & $(0,\,2)$ & $\mathrm{I}\text{-}01$ & $(0,\,0,\,2)$ & $(0,\,0,\,1)$ \\
		& $(3,\,2)$ & $\mathrm{I}\text{-}00$ & $(1,\,2,\,0)$ & $(0,\,0,\,2)$ &  & $(1,\,2)$ & $\mathrm{I}\text{-}11$ & $(0,\,0,\,1)$ & $(0,\,0,\,1)$ \\
		$(1,\,1,\,0)$ & $(0,\,1)$ & $\mathrm{I}\text{-}00$ & $(0,\,0,\,1)$ & $(1,\,0,\,0)$ &  & $(2,\,2)$ & $\mathrm{I}\text{-}01$ & $(1,\,1,\,1)$ & $(0,\,0,\,1)$ \\
		& $(1,\,1)$ & $\mathrm{I}\text{-}10$ & $-$ & $(1,\,0,\,0)$ &  & $(3,\,2)$ & $\mathrm{I}\text{-}11$ & $(1,\,1,\,0)$ & $(0,\,0,\,1)$ \\
		& $(2,\,1)$ & $\mathrm{I}\text{-}00$ & $(1,\,1,\,0)$ & $(1,\,0,\,0)$ &  & $(4,\,2)$ & $\mathrm{I}\text{-}01$ & $(2,\,2,\,0)$ & $(0,\,0,\,1)$ \\
		$(1,\,1,\,1)$ & $(1,\,0)$ & $\mathrm{I}\text{-}00$ & $(1,\,0,\,0)$ & $(0,\,1,\,0)$ &  & $(1,\,3)$ & $\mathrm{I}\text{-}00$ & $(0,\,1,\,2)$ & $(1,\,0,\,2)$ \\
		& $(0,\,1)$ & $\mathrm{I}\text{-}01$ & $(0,\,0,\,1)$ & $-$ &  & $(2,\,3)$ & $\mathrm{I}\text{-}10$ & $(0,\,1,\,1)$ & $(1,\,0,\,2)$ \\
		& $(1,\,1)$ & $\mathrm{I}\text{-}11$ & $-$ & $-$ &  & $(3,\,3)$ & $\mathrm{I}\text{-}00$ & $(1,\,2,\,1)$ & $(1,\,0,\,2)$ \\
		& $(2,\,1)$ & $\mathrm{I}\text{-}01$ & $(1,\,1,\,0)$ & $-$ &  & $(4,\,3)$ & $\mathrm{I}\text{-}10$ & $(1,\,2,\,0)$ & $(1,\,0,\,2)$ \\
		& $(1,\,2)$ & $\mathrm{I}\text{-}00$ & $(0,\,1,\,1)$ & $(1,\,0,\,1)$ &  & $(5,\,3)$ & $\mathrm{I}\text{-}00$ & $(2,\,3,\,0)$ & $(1,\,0,\,2)$ \\
		& $(2,\,2)$ & $\mathrm{I}\text{-}10$ & $(0,\,1,\,0)$ & $(1,\,0,\,1)$ & $(2,\,2,\,0)$ & $(0,\,2)$ & $\mathrm{I}\text{-}00$ & $(0,\,0,\,2)$ & $(2,\,0,\,0)$ \\
		& $(3,\,2)$ & $\mathrm{I}\text{-}00$ & $(1,\,2,\,0)$ & $(1,\,0,\,1)$ &  & $(1,\,2)$ & $\mathrm{I}\text{-}10$ & $(0,\,0,\,1)$ & $(2,\,0,\,0)$ \\
		$(1,\,1,\,2)$ & $(0,\,1)$ & $\mathrm{I}\text{-}00$ & $(0,\,0,\,1)$ & $(0,\,1,\,1)$ &  & $(2,\,2)$ & $\mathrm{I}\text{-}00$ & $(1,\,1,\,1)$ & $(2,\,0,\,0)$ \\
		& $(1,\,1)$ & $\mathrm{I}\text{-}10$ & $-$ & $(0,\,1,\,1)$ &  & $(3,\,2)$ & $\mathrm{I}\text{-}10$ & $(1,\,1,\,0)$ & $(2,\,0,\,0)$ \\
		& $(2,\,1)$ & $\mathrm{I}\text{-}00$ & $(1,\,1,\,0)$ & $(0,\,1,\,1)$ &  & $(4,\,2)$ & $\mathrm{I}\text{-}00$ & $(2,\,2,\,0)$ & $(2,\,0,\,0)$ \\
		& $(1,\,2)$ & $\mathrm{I}\text{-}01$ & $(0,\,1,\,1)$ & $(0,\,0,\,1)$ & $(2,\,2,\,1)$ & $(1,\,1)$ & $\mathrm{I}\text{-}00$ & $(1,\,0,\,1)$ & $(1,\,1,\,0)$ \\
		& $(2,\,2)$ & $\mathrm{I}\text{-}11$ & $(0,\,1,\,0)$ & $(0,\,0,\,1)$ &  & $(2,\,1)$ & $\mathrm{I}\text{-}10$ & $(1,\,0,\,0)$ & $(1,\,1,\,0)$ \\
		& $(3,\,2)$ & $\mathrm{I}\text{-}01$ & $(1,\,2,\,0)$ & $(0,\,0,\,1)$ &  & $(3,\,1)$ & $\mathrm{I}\text{-}00$ & $(2,\,1,\,0)$ & $(1,\,1,\,0)$ \\
		& $(2,\,3)$ & $\mathrm{I}\text{-}00$ & $(0,\,2,\,1)$ & $(1,\,0,\,2)$ &  & $(0,\,2)$ & $\mathrm{I}\text{-}01$ & $(0,\,0,\,2)$ & $(1,\,0,\,0)$ \\
		& $(3,\,3)$ & $\mathrm{I}\text{-}10$ & $(0,\,2,\,0)$ & $(1,\,0,\,2)$ &  & $(1,\,2)$ & $\mathrm{I}\text{-}11$ & $(0,\,0,\,1)$ & $(1,\,0,\,0)$ \\
		& $(4,\,3)$ & $\mathrm{I}\text{-}00$ & $(1,\,3,\,0)$ & $(1,\,0,\,2)$ &  & $(2,\,2)$ & $\mathrm{I}\text{-}01$ & $(1,\,1,\,1)$ & $(1,\,0,\,0)$ \\
		$(1,\,2,\,0)$ & $(1,\,2)$ & $\mathrm{I}\text{-}00$ & $(0,\,1,\,1)$ & $(2,\,0,\,0)$ &  & $(3,\,2)$ & $\mathrm{I}\text{-}11$ & $(1,\,1,\,0)$ & $(1,\,0,\,0)$ \\
		& $(2,\,2)$ & $\mathrm{I}\text{-}10$ & $(0,\,1,\,0)$ & $(2,\,0,\,0)$ &  & $(4,\,2)$ & $\mathrm{I}\text{-}01$ & $(2,\,2,\,0)$ & $(1,\,0,\,0)$ \\
		& $(3,\,2)$ & $\mathrm{I}\text{-}00$ & $(1,\,2,\,0)$ & $(2,\,0,\,0)$ &  & $(1,\,3)$ & $\mathrm{I}\text{-}00$ & $(0,\,1,\,2)$ & $(2,\,0,\,1)$ \\
		$(1,\,2,\,1)$ & $(0,\,1)$ & $\mathrm{I}\text{-}00$ & $(0,\,0,\,1)$ & $(1,\,1,\,0)$ &  & $(2,\,3)$ & $\mathrm{I}\text{-}10$ & $(0,\,1,\,1)$ & $(2,\,0,\,1)$ \\
		& $(1,\,1)$ & $\mathrm{I}\text{-}10$ & $-$ & $(1,\,1,\,0)$ &  & $(3,\,3)$ & $\mathrm{I}\text{-}00$ & $(1,\,2,\,1)$ & $(2,\,0,\,1)$ \\
		& $(2,\,1)$ & $\mathrm{I}\text{-}00$ & $(1,\,1,\,0)$ & $(1,\,1,\,0)$ &  & $(4,\,3)$ & $\mathrm{I}\text{-}10$ & $(1,\,2,\,0)$ & $(2,\,0,\,1)$ \\
		& $(1,\,2)$ & $\mathrm{I}\text{-}01$ & $(0,\,1,\,1)$ & $(1,\,0,\,0)$ &  & $(5,\,3)$ & $\mathrm{I}\text{-}00$ & $(2,\,3,\,0)$ & $(2,\,0,\,1)$ \\
		& $(2,\,2)$ & $\mathrm{I}\text{-}11$ & $(0,\,1,\,0)$ & $(1,\,0,\,0)$ & $(2,\,2,\,2)$ & $(2,\,0)$ & $\mathrm{I}\text{-}00$ & $(2,\,0,\,0)$ & $(0,\,2,\,0)$ \\
		& $(3,\,2)$ & $\mathrm{I}\text{-}01$ & $(1,\,2,\,0)$ & $(1,\,0,\,0)$ &  & $(1,\,1)$ & $\mathrm{I}\text{-}01$ & $(1,\,0,\,1)$ & $(0,\,1,\,0)$ \\
		& $(2,\,3)$ & $\mathrm{I}\text{-}00$ & $(0,\,2,\,1)$ & $(2,\,0,\,1)$ &  & $(2,\,1)$ & $\mathrm{I}\text{-}11$ & $(1,\,0,\,0)$ & $(0,\,1,\,0)$ \\
		& $(3,\,3)$ & $\mathrm{I}\text{-}10$ & $(0,\,2,\,0)$ & $(2,\,0,\,1)$ &  & $(3,\,1)$ & $\mathrm{I}\text{-}01$ & $(2,\,1,\,0)$ & $(0,\,1,\,0)$ \\
		& $(4,\,3)$ & $\mathrm{I}\text{-}00$ & $(1,\,3,\,0)$ & $(2,\,0,\,1)$ &  & $(0,\,2)$ & $\mathrm{I}\text{-}00$ & $(0,\,0,\,2)$ & $(1,\,1,\,1)$ \\
		$(1,\,2,\,2)$ & $(1,\,0)$ & $\mathrm{I}\text{-}00$ & $(1,\,0,\,0)$ & $(0,\,2,\,0)$ &  & $(1,\,2)$ & $\mathrm{I}\text{-}10$ & $(0,\,0,\,1)$ & $(1,\,1,\,1)$ \\
		& $(0,\,1)$ & $\mathrm{I}\text{-}01$ & $(0,\,0,\,1)$ & $(0,\,1,\,0)$ &  & $(2,\,2)$ & $\mathrm{I}\text{-}00$ & $(1,\,1,\,1)$ & $(1,\,1,\,1)$ \\
		& $(1,\,1)$ & $\mathrm{I}\text{-}11$ & $-$ & $(0,\,1,\,0)$ &  & $(3,\,2)$ & $\mathrm{I}\text{-}10$ & $(1,\,1,\,0)$ & $(1,\,1,\,1)$ \\
		& $(2,\,1)$ & $\mathrm{I}\text{-}01$ & $(1,\,1,\,0)$ & $(0,\,1,\,0)$ &  & $(4,\,2)$ & $\mathrm{I}\text{-}00$ & $(2,\,2,\,0)$ & $(1,\,1,\,1)$ \\
		& $(1,\,2)$ & $\mathrm{I}\text{-}00$ & $(0,\,1,\,1)$ & $(1,\,1,\,1)$ &  & $(1,\,3)$ & $\mathrm{I}\text{-}01$ & $(0,\,1,\,2)$ & $(1,\,0,\,1)$ \\
		& $(2,\,2)$ & $\mathrm{I}\text{-}10$ & $(0,\,1,\,0)$ & $(1,\,1,\,1)$ &  & $(2,\,3)$ & $\mathrm{I}\text{-}11$ & $(0,\,1,\,1)$ & $(1,\,0,\,1)$ \\
		& $(3,\,2)$ & $\mathrm{I}\text{-}00$ & $(1,\,2,\,0)$ & $(1,\,1,\,1)$ &  & $(3,\,3)$ & $\mathrm{I}\text{-}01$ & $(1,\,2,\,1)$ & $(1,\,0,\,1)$ \\
		& $(2,\,3)$ & $\mathrm{I}\text{-}01$ & $(0,\,2,\,1)$ & $(1,\,0,\,1)$ &  & $(4,\,3)$ & $\mathrm{I}\text{-}11$ & $(1,\,2,\,0)$ & $(1,\,0,\,1)$ \\
		& $(3,\,3)$ & $\mathrm{I}\text{-}11$ & $(0,\,2,\,0)$ & $(1,\,0,\,1)$ &  & $(5,\,3)$ & $\mathrm{I}\text{-}01$ & $(2,\,3,\,0)$ & $(1,\,0,\,1)$ \\
		& $(4,\,3)$ & $\mathrm{I}\text{-}01$ & $(1,\,3,\,0)$ & $(1,\,0,\,1)$ &  & $(2,\,4)$ & $\mathrm{I}\text{-}00$ & $(0,\,2,\,2)$ & $(2,\,0,\,2)$ \\
		& $(3,\,4)$ & $\mathrm{I}\text{-}00$ & $(0,\,3,\,1)$ & $(2,\,0,\,2)$ &  & $(3,\,4)$ & $\mathrm{I}\text{-}10$ & $(0,\,2,\,1)$ & $(2,\,0,\,2)$ \\
		& $(4,\,4)$ & $\mathrm{I}\text{-}10$ & $(0,\,3,\,0)$ & $(2,\,0,\,2)$ &  & $(4,\,4)$ & $\mathrm{I}\text{-}00$ & $(1,\,3,\,1)$ & $(2,\,0,\,2)$ \\
		& $(5,\,4)$ & $\mathrm{I}\text{-}00$ & $(1,\,4,\,0)$ & $(2,\,0,\,2)$ &  & $(5,\,4)$ & $\mathrm{I}\text{-}10$ & $(1,\,3,\,0)$ & $(2,\,0,\,2)$ \\
		&  &  &  &  &  & $(6,\,4)$ & $\mathrm{I}\text{-}00$ & $(2,\,4,\,0)$ & $(2,\,0,\,2)$ \\
		\hline
		\multicolumn{10}{c}{\textbf{Case II: $s_1$ integer spin, $s_2,s_3$ half-integer spins}} \\
		\hline
		$(0,\,\tfrac12,\,\tfrac12)$ & $(0,\,0)$ & $\mathrm{II}\text{-}01$ & $-$ & $-$ & $(2,\,\tfrac12,\,\tfrac12)$ & $(2,\,0)$ & $\mathrm{II}\text{-}01$ & $(2,\,0,\,0)$ & $-$ \\
		& $(1,\,1)$ & $\mathrm{II}\text{-}00$ & $(0,\,1,\,0)$ & $-$ &  & $(1,\,1)$ & $\mathrm{II}\text{-}00$ & $(1,\,0,\,1)$ & $-$ \\
		$(0,\,\tfrac12,\,\tfrac32)$ & $(1,\,1)$ & $\mathrm{II}\text{-}01$ & $(0,\,1,\,0)$ & $(0,\,0,\,1)$ &  & $(2,\,1)$ & $\mathrm{II}\text{-}10$ & $(1,\,0,\,0)$ & $-$ \\
		& $(2,\,2)$ & $\mathrm{II}\text{-}00$ & $(0,\,2,\,0)$ & $(0,\,0,\,1)$ &  & $(3,\,1)$ & $\mathrm{II}\text{-}00$ & $(2,\,1,\,0)$ & $-$ \\
		$(0,\,\tfrac32,\,\tfrac12)$ & $(1,\,1)$ & $\mathrm{II}\text{-}01$ & $(0,\,1,\,0)$ & $(1,\,0,\,0)$ & $(2,\,\tfrac12,\,\tfrac32)$ & $(1,\,1)$ & $\mathrm{II}\text{-}01$ & $(1,\,0,\,1)$ & $(0,\,0,\,1)$ \\
		& $(2,\,2)$ & $\mathrm{II}\text{-}00$ & $(0,\,2,\,0)$ & $(1,\,0,\,0)$ &  & $(2,\,1)$ & $\mathrm{II}\text{-}11$ & $(1,\,0,\,0)$ & $(0,\,0,\,1)$ \\
		$(0,\,\tfrac32,\,\tfrac32)$ & $(0,\,0)$ & $\mathrm{II}\text{-}01$ & $-$ & $(0,\,1,\,0)$ &  & $(3,\,1)$ & $\mathrm{II}\text{-}01$ & $(2,\,1,\,0)$ & $(0,\,0,\,1)$ \\
		& $(1,\,1)$ & $\mathrm{II}\text{-}00$ & $(0,\,1,\,0)$ & $(0,\,1,\,0)$ &  & $(0,\,2)$ & $\mathrm{II}\text{-}00$ & $(0,\,0,\,2)$ & $(0,\,0,\,1)$ \\
		& $(2,\,2)$ & $\mathrm{II}\text{-}01$ & $(0,\,2,\,0)$ & $(1,\,0,\,1)$ &  & $(1,\,2)$ & $\mathrm{II}\text{-}10$ & $(0,\,0,\,1)$ & $(0,\,0,\,1)$ \\
		& $(3,\,3)$ & $\mathrm{II}\text{-}00$ & $(0,\,3,\,0)$ & $(1,\,0,\,1)$ &  & $(2,\,2)$ & $\mathrm{II}\text{-}00$ & $(1,\,1,\,1)$ & $(0,\,0,\,1)$ \\
		$(1,\,\tfrac12,\,\tfrac12)$ & $(1,\,0)$ & $\mathrm{II}\text{-}01$ & $(1,\,0,\,0)$ & $-$ &  & $(3,\,2)$ & $\mathrm{II}\text{-}10$ & $(1,\,1,\,0)$ & $(0,\,0,\,1)$ \\
		& $(0,\,1)$ & $\mathrm{II}\text{-}00$ & $(0,\,0,\,1)$ & $-$ &  & $(4,\,2)$ & $\mathrm{II}\text{-}00$ & $(2,\,2,\,0)$ & $(0,\,0,\,1)$ \\
		& $(1,\,1)$ & $\mathrm{II}\text{-}10$ & $-$ & $-$ & $(2,\,\tfrac32,\,\tfrac12)$ & $(1,\,1)$ & $\mathrm{II}\text{-}01$ & $(1,\,0,\,1)$ & $(1,\,0,\,0)$ \\
		& $(2,\,1)$ & $\mathrm{II}\text{-}00$ & $(1,\,1,\,0)$ & $-$ &  & $(2,\,1)$ & $\mathrm{II}\text{-}11$ & $(1,\,0,\,0)$ & $(1,\,0,\,0)$ \\
		$(1,\,\tfrac12,\,\tfrac32)$ & $(0,\,1)$ & $\mathrm{II}\text{-}01$ & $(0,\,0,\,1)$ & $(0,\,0,\,1)$ &  & $(3,\,1)$ & $\mathrm{II}\text{-}01$ & $(2,\,1,\,0)$ & $(1,\,0,\,0)$ \\
		& $(1,\,1)$ & $\mathrm{II}\text{-}11$ & $-$ & $(0,\,0,\,1)$ &  & $(0,\,2)$ & $\mathrm{II}\text{-}00$ & $(0,\,0,\,2)$ & $(1,\,0,\,0)$ \\
		& $(2,\,1)$ & $\mathrm{II}\text{-}01$ & $(1,\,1,\,0)$ & $(0,\,0,\,1)$ &  & $(1,\,2)$ & $\mathrm{II}\text{-}10$ & $(0,\,0,\,1)$ & $(1,\,0,\,0)$ \\
		& $(1,\,2)$ & $\mathrm{II}\text{-}00$ & $(0,\,1,\,1)$ & $(0,\,0,\,1)$ &  & $(2,\,2)$ & $\mathrm{II}\text{-}00$ & $(1,\,1,\,1)$ & $(1,\,0,\,0)$ \\
		& $(2,\,2)$ & $\mathrm{II}\text{-}10$ & $(0,\,1,\,0)$ & $(0,\,0,\,1)$ &  & $(3,\,2)$ & $\mathrm{II}\text{-}10$ & $(1,\,1,\,0)$ & $(1,\,0,\,0)$ \\
		& $(3,\,2)$ & $\mathrm{II}\text{-}00$ & $(1,\,2,\,0)$ & $(0,\,0,\,1)$ &  & $(4,\,2)$ & $\mathrm{II}\text{-}00$ & $(2,\,2,\,0)$ & $(1,\,0,\,0)$ \\
		$(1,\,\tfrac32,\,\tfrac12)$ & $(0,\,1)$ & $\mathrm{II}\text{-}01$ & $(0,\,0,\,1)$ & $(1,\,0,\,0)$ & $(2,\,\tfrac32,\,\tfrac32)$ & $(2,\,0)$ & $\mathrm{II}\text{-}01$ & $(2,\,0,\,0)$ & $(0,\,1,\,0)$ \\
		& $(1,\,1)$ & $\mathrm{II}\text{-}11$ & $-$ & $(1,\,0,\,0)$ &  & $(1,\,1)$ & $\mathrm{II}\text{-}00$ & $(1,\,0,\,1)$ & $(0,\,1,\,0)$ \\
		& $(2,\,1)$ & $\mathrm{II}\text{-}01$ & $(1,\,1,\,0)$ & $(1,\,0,\,0)$ &  & $(2,\,1)$ & $\mathrm{II}\text{-}10$ & $(1,\,0,\,0)$ & $(0,\,1,\,0)$ \\
		& $(1,\,2)$ & $\mathrm{II}\text{-}00$ & $(0,\,1,\,1)$ & $(1,\,0,\,0)$ &  & $(3,\,1)$ & $\mathrm{II}\text{-}00$ & $(2,\,1,\,0)$ & $(0,\,1,\,0)$ \\
		& $(2,\,2)$ & $\mathrm{II}\text{-}10$ & $(0,\,1,\,0)$ & $(1,\,0,\,0)$ &  & $(0,\,2)$ & $\mathrm{II}\text{-}01$ & $(0,\,0,\,2)$ & $(1,\,0,\,1)$ \\
		& $(3,\,2)$ & $\mathrm{II}\text{-}00$ & $(1,\,2,\,0)$ & $(1,\,0,\,0)$ &  & $(1,\,2)$ & $\mathrm{II}\text{-}11$ & $(0,\,0,\,1)$ & $(1,\,0,\,1)$ \\
		$(1,\,\tfrac32,\,\tfrac32)$ & $(1,\,0)$ & $\mathrm{II}\text{-}01$ & $(1,\,0,\,0)$ & $(0,\,1,\,0)$ &  & $(2,\,2)$ & $\mathrm{II}\text{-}01$ & $(1,\,1,\,1)$ & $(1,\,0,\,1)$ \\
		& $(0,\,1)$ & $\mathrm{II}\text{-}00$ & $(0,\,0,\,1)$ & $(0,\,1,\,0)$ &  & $(3,\,2)$ & $\mathrm{II}\text{-}11$ & $(1,\,1,\,0)$ & $(1,\,0,\,1)$ \\
		& $(1,\,1)$ & $\mathrm{II}\text{-}10$ & $-$ & $(0,\,1,\,0)$ &  & $(4,\,2)$ & $\mathrm{II}\text{-}01$ & $(2,\,2,\,0)$ & $(1,\,0,\,1)$ \\
		& $(2,\,1)$ & $\mathrm{II}\text{-}00$ & $(1,\,1,\,0)$ & $(0,\,1,\,0)$ &  & $(1,\,3)$ & $\mathrm{II}\text{-}00$ & $(0,\,1,\,2)$ & $(1,\,0,\,1)$ \\
		& $(1,\,2)$ & $\mathrm{II}\text{-}01$ & $(0,\,1,\,1)$ & $(1,\,0,\,1)$ &  & $(2,\,3)$ & $\mathrm{II}\text{-}10$ & $(0,\,1,\,1)$ & $(1,\,0,\,1)$ \\
		& $(2,\,2)$ & $\mathrm{II}\text{-}11$ & $(0,\,1,\,0)$ & $(1,\,0,\,1)$ &  & $(3,\,3)$ & $\mathrm{II}\text{-}00$ & $(1,\,2,\,1)$ & $(1,\,0,\,1)$ \\
		& $(3,\,2)$ & $\mathrm{II}\text{-}01$ & $(1,\,2,\,0)$ & $(1,\,0,\,1)$ &  & $(4,\,3)$ & $\mathrm{II}\text{-}10$ & $(1,\,2,\,0)$ & $(1,\,0,\,1)$ \\
		& $(2,\,3)$ & $\mathrm{II}\text{-}00$ & $(0,\,2,\,1)$ & $(1,\,0,\,1)$ &  & $(5,\,3)$ & $\mathrm{II}\text{-}00$ & $(2,\,3,\,0)$ & $(1,\,0,\,1)$ \\
		& $(3,\,3)$ & $\mathrm{II}\text{-}10$ & $(0,\,2,\,0)$ & $(1,\,0,\,1)$ &  &  &  &  &  \\
		& $(4,\,3)$ & $\mathrm{II}\text{-}00$ & $(1,\,3,\,0)$ & $(1,\,0,\,1)$ &  &  &  &  &  \\
		\hline
		\multicolumn{10}{c}{\textbf{Case III: $s_1,s_3$ half-integer spins, $s_2$ integer spin}} \\
		\hline
		$(\tfrac12,\,0,\,\tfrac12)$ & $(0,\,\tfrac12)$ & $\mathrm{III}\text{-}11$ & $-$ & $-$ & $(\tfrac32,\,1,\,\tfrac12)$ & $(1,\,\tfrac12)$ & $\mathrm{III}\text{-}10$ & $(1,\,0,\,0)$ & $-$ \\
		& $(1,\,\tfrac12)$ & $\mathrm{III}\text{-}01$ & $-$ & $-$ &  & $(2,\,\tfrac12)$ & $\mathrm{III}\text{-}00$ & $(1,\,0,\,0)$ & $-$ \\
		$(\tfrac12,\,0,\,\tfrac32)$ & $(1,\,\tfrac32)$ & $\mathrm{III}\text{-}11$ & $(0,\,1,\,0)$ & $(0,\,0,\,1)$ &  & $(0,\,\tfrac32)$ & $\mathrm{III}\text{-}11$ & $(0,\,0,\,1)$ & $(1,\,0,\,0)$ \\
		& $(2,\,\tfrac32)$ & $\mathrm{III}\text{-}01$ & $(0,\,1,\,0)$ & $(0,\,0,\,1)$ &  & $(1,\,\tfrac32)$ & $\mathrm{III}\text{-}01$ & $(0,\,0,\,1)$ & $(1,\,0,\,0)$ \\
		$(\tfrac12,\,1,\,\tfrac12)$ & $(0,\,\tfrac12)$ & $\mathrm{III}\text{-}10$ & $-$ & $-$ &  & $(2,\,\tfrac32)$ & $\mathrm{III}\text{-}11$ & $(1,\,1,\,0)$ & $(1,\,0,\,0)$ \\
		& $(1,\,\tfrac12)$ & $\mathrm{III}\text{-}00$ & $-$ & $-$ &  & $(3,\,\tfrac32)$ & $\mathrm{III}\text{-}01$ & $(1,\,1,\,0)$ & $(1,\,0,\,0)$ \\
		& $(1,\,\tfrac32)$ & $\mathrm{III}\text{-}11$ & $(0,\,1,\,0)$ & $(1,\,0,\,0)$ & $(\tfrac32,\,1,\,\tfrac32)$ & $(1,\,\tfrac12)$ & $\mathrm{III}\text{-}11$ & $(1,\,0,\,0)$ & $(0,\,1,\,0)$ \\
		& $(2,\,\tfrac32)$ & $\mathrm{III}\text{-}01$ & $(0,\,1,\,0)$ & $(1,\,0,\,0)$ &  & $(2,\,\tfrac12)$ & $\mathrm{III}\text{-}01$ & $(1,\,0,\,0)$ & $(0,\,1,\,0)$ \\
		$(\tfrac12,\,1,\,\tfrac32)$ & $(0,\,\tfrac12)$ & $\mathrm{III}\text{-}11$ & $-$ & $(0,\,1,\,0)$ &  & $(0,\,\tfrac32)$ & $\mathrm{III}\text{-}10$ & $(0,\,0,\,1)$ & $(0,\,0,\,1)$ \\
		& $(1,\,\tfrac12)$ & $\mathrm{III}\text{-}01$ & $-$ & $(0,\,1,\,0)$ &  & $(1,\,\tfrac32)$ & $\mathrm{III}\text{-}00$ & $(0,\,0,\,1)$ & $(0,\,0,\,1)$ \\
		& $(1,\,\tfrac32)$ & $\mathrm{III}\text{-}10$ & $(0,\,1,\,0)$ & $(0,\,0,\,1)$ &  & $(2,\,\tfrac32)$ & $\mathrm{III}\text{-}10$ & $(1,\,1,\,0)$ & $(0,\,0,\,1)$ \\
		& $(2,\,\tfrac32)$ & $\mathrm{III}\text{-}00$ & $(0,\,1,\,0)$ & $(0,\,0,\,1)$ &  & $(3,\,\tfrac32)$ & $\mathrm{III}\text{-}00$ & $(1,\,1,\,0)$ & $(0,\,0,\,1)$ \\
		& $(2,\,\tfrac52)$ & $\mathrm{III}\text{-}11$ & $(0,\,2,\,0)$ & $(1,\,0,\,1)$ &  & $(1,\,\tfrac52)$ & $\mathrm{III}\text{-}11$ & $(0,\,1,\,1)$ & $(1,\,0,\,1)$ \\
		& $(3,\,\tfrac52)$ & $\mathrm{III}\text{-}01$ & $(0,\,2,\,0)$ & $(1,\,0,\,1)$ &  & $(2,\,\tfrac52)$ & $\mathrm{III}\text{-}01$ & $(0,\,1,\,1)$ & $(1,\,0,\,1)$ \\
		$(\tfrac12,\,2,\,\tfrac12)$ & $(1,\,\tfrac32)$ & $\mathrm{III}\text{-}10$ & $(0,\,1,\,0)$ & $(1,\,0,\,0)$ &  & $(3,\,\tfrac52)$ & $\mathrm{III}\text{-}11$ & $(1,\,2,\,0)$ & $(1,\,0,\,1)$ \\
		& $(2,\,\tfrac32)$ & $\mathrm{III}\text{-}00$ & $(0,\,1,\,0)$ & $(1,\,0,\,0)$ &  & $(4,\,\tfrac52)$ & $\mathrm{III}\text{-}01$ & $(1,\,2,\,0)$ & $(1,\,0,\,1)$ \\
		& $(2,\,\tfrac52)$ & $\mathrm{III}\text{-}11$ & $(0,\,2,\,0)$ & $(2,\,0,\,0)$ & $(\tfrac32,\,2,\,\tfrac12)$ & $(0,\,\tfrac32)$ & $\mathrm{III}\text{-}10$ & $(0,\,0,\,1)$ & $(1,\,0,\,0)$ \\
		& $(3,\,\tfrac52)$ & $\mathrm{III}\text{-}01$ & $(0,\,2,\,0)$ & $(2,\,0,\,0)$ &  & $(1,\,\tfrac32)$ & $\mathrm{III}\text{-}00$ & $(0,\,0,\,1)$ & $(1,\,0,\,0)$ \\
		$(\tfrac12,\,2,\,\tfrac32)$ & $(0,\,\tfrac12)$ & $\mathrm{III}\text{-}10$ & $-$ & $(0,\,1,\,0)$ &  & $(2,\,\tfrac32)$ & $\mathrm{III}\text{-}10$ & $(1,\,1,\,0)$ & $(1,\,0,\,0)$ \\
		& $(1,\,\tfrac12)$ & $\mathrm{III}\text{-}00$ & $-$ & $(0,\,1,\,0)$ &  & $(3,\,\tfrac32)$ & $\mathrm{III}\text{-}00$ & $(1,\,1,\,0)$ & $(1,\,0,\,0)$ \\
		& $(1,\,\tfrac32)$ & $\mathrm{III}\text{-}11$ & $(0,\,1,\,0)$ & $(1,\,1,\,0)$ &  & $(1,\,\tfrac52)$ & $\mathrm{III}\text{-}11$ & $(0,\,1,\,1)$ & $(2,\,0,\,0)$ \\
		& $(2,\,\tfrac32)$ & $\mathrm{III}\text{-}01$ & $(0,\,1,\,0)$ & $(1,\,1,\,0)$ &  & $(2,\,\tfrac52)$ & $\mathrm{III}\text{-}01$ & $(0,\,1,\,1)$ & $(2,\,0,\,0)$ \\
		& $(2,\,\tfrac52)$ & $\mathrm{III}\text{-}10$ & $(0,\,2,\,0)$ & $(1,\,0,\,1)$ &  & $(3,\,\tfrac52)$ & $\mathrm{III}\text{-}11$ & $(1,\,2,\,0)$ & $(2,\,0,\,0)$ \\
		& $(3,\,\tfrac52)$ & $\mathrm{III}\text{-}00$ & $(0,\,2,\,0)$ & $(1,\,0,\,1)$ &  & $(4,\,\tfrac52)$ & $\mathrm{III}\text{-}01$ & $(1,\,2,\,0)$ & $(2,\,0,\,0)$ \\
		& $(3,\,\tfrac72)$ & $\mathrm{III}\text{-}11$ & $(0,\,3,\,0)$ & $(2,\,0,\,1)$ & $(\tfrac32,\,2,\,\tfrac32)$ & $(1,\,\tfrac12)$ & $\mathrm{III}\text{-}10$ & $(1,\,0,\,0)$ & $(0,\,1,\,0)$ \\
		& $(4,\,\tfrac72)$ & $\mathrm{III}\text{-}01$ & $(0,\,3,\,0)$ & $(2,\,0,\,1)$ &  & $(2,\,\tfrac12)$ & $\mathrm{III}\text{-}00$ & $(1,\,0,\,0)$ & $(0,\,1,\,0)$ \\
		$(\tfrac32,\,0,\,\tfrac12)$ & $(1,\,\tfrac12)$ & $\mathrm{III}\text{-}11$ & $(1,\,0,\,0)$ & $-$ &  & $(0,\,\tfrac32)$ & $\mathrm{III}\text{-}11$ & $(0,\,0,\,1)$ & $(1,\,1,\,0)$ \\
		& $(2,\,\tfrac12)$ & $\mathrm{III}\text{-}01$ & $(1,\,0,\,0)$ & $-$ &  & $(1,\,\tfrac32)$ & $\mathrm{III}\text{-}01$ & $(0,\,0,\,1)$ & $(1,\,1,\,0)$ \\
		$(\tfrac32,\,0,\,\tfrac32)$ & $(0,\,\tfrac32)$ & $\mathrm{III}\text{-}11$ & $(0,\,0,\,1)$ & $(0,\,0,\,1)$ &  & $(2,\,\tfrac32)$ & $\mathrm{III}\text{-}11$ & $(1,\,1,\,0)$ & $(1,\,1,\,0)$ \\
		& $(1,\,\tfrac32)$ & $\mathrm{III}\text{-}01$ & $(0,\,0,\,1)$ & $(0,\,0,\,1)$ &  & $(3,\,\tfrac32)$ & $\mathrm{III}\text{-}01$ & $(1,\,1,\,0)$ & $(1,\,1,\,0)$ \\
		& $(2,\,\tfrac32)$ & $\mathrm{III}\text{-}11$ & $(1,\,1,\,0)$ & $(0,\,0,\,1)$ &  & $(1,\,\tfrac52)$ & $\mathrm{III}\text{-}10$ & $(0,\,1,\,1)$ & $(1,\,0,\,1)$ \\
		& $(3,\,\tfrac32)$ & $\mathrm{III}\text{-}01$ & $(1,\,1,\,0)$ & $(0,\,0,\,1)$ &  & $(2,\,\tfrac52)$ & $\mathrm{III}\text{-}00$ & $(0,\,1,\,1)$ & $(1,\,0,\,1)$ \\
		&  &  &  &  &  & $(3,\,\tfrac52)$ & $\mathrm{III}\text{-}10$ & $(1,\,2,\,0)$ & $(1,\,0,\,1)$ \\
		&  &  &  &  &  & $(4,\,\tfrac52)$ & $\mathrm{III}\text{-}00$ & $(1,\,2,\,0)$ & $(1,\,0,\,1)$ \\
		&  &  &  &  &  & $(2,\,\tfrac72)$ & $\mathrm{III}\text{-}11$ & $(0,\,2,\,1)$ & $(2,\,0,\,1)$ \\
		&  &  &  &  &  & $(3,\,\tfrac72)$ & $\mathrm{III}\text{-}01$ & $(0,\,2,\,1)$ & $(2,\,0,\,1)$ \\
		&  &  &  &  &  & $(4,\,\tfrac72)$ & $\mathrm{III}\text{-}11$ & $(1,\,3,\,0)$ & $(2,\,0,\,1)$ \\
		&  &  &  &  &  & $(5,\,\tfrac72)$ & $\mathrm{III}\text{-}01$ & $(1,\,3,\,0)$ & $(2,\,0,\,1)$ \\
	\end{longtable}
\end{footnotesize}

	\subsection*{B. Explicit examples of the covariant on-shell partial-wave amplitudes}

In this part we illustrate the construction procedure of the covariant on-shell $\mathcal{H}$-type partial-wave amplitudes with five explicit examples.
For each example we (i) determine the angular-momentum-allowed $L$-$S$ combinations from the spin quantum numbers, (ii) read off the amplitude type $\mathrm{X}\text{-}\delta_1\delta_2$ and the numbers of indices in the index sets from the table of index sets given above, (iii) write down the amplitude from the formulas in Table~II of the main text with the corresponding index sets, and (iv) simplify the result in the rest frame of the parent particle using the properties of the spin projection tensors and of the on-shell spin wave functions, and compare with the results in the literature.
The comparison is made with the explicit formulas of the covariant tensor formalism of Ref.~\cite{Zou:2002ar} and of the covariant $L$-$S$ scheme of Ref.~\cite{Zou:2002yy}.

\textbf{Simplification identities.}
In the rest frame of the parent particle, $v_1=(1,\bm{0})$, the spin projection tensor of spin $1$ reads
\begin{equation}
	P^{(1)}_{\mu\nu}(p_1)=\tilde{g}_{\mu\nu}(p_1)=g_{\mu\nu}-v_{1\mu}v_{1\nu},\qquad
	P^{(1)}_{\mu}{}^{\alpha}(p_1)\,P^{(1)}_{\alpha\nu}(p_1)=P^{(1)}_{\mu\nu}(p_1),
	\label{sm:P1-idempotent}
\end{equation}
i.e., it is idempotent, $P^{(1)}P^{(1)}=P^{(1)}$, as a projection operator must be; the same normalization is used for the higher-spin projection tensors, e.g., $P^{(3/2)}_{\mu\nu}=\tilde{g}_{\mu\nu}-\frac{1}{3}\tilde{\gamma}_\mu\tilde{\gamma}_\nu$ with $\tilde{\gamma}_\mu=P^{(1)}_{\mu\nu}\gamma^\nu$, which obeys $P^{(3/2)}P^{(3/2)}=P^{(3/2)}$ as well because $\tilde{\gamma}\cdot\tilde{\gamma}=3$.
The opposite sign, $P^{(1)}_{\mu\nu}=-g_{\mu\nu}+v_{1\mu}v_{1\nu}$, is also used in the literature (e.g., in Ref.~\cite{Zou:2002ar}), in which case $P^{(1)}P^{(1)}=-P^{(1)}$; the two conventions differ by an overall sign that is absorbed into the arbitrary normalization of the coupling tensors (see the discussion of coupling-tensor normalization
in the main text).
The orbital angular momentum tensor of rank one is the $p_1$-transversalization of the relative momentum, i.e., $q$ with its component along $v_1$ removed,
\begin{equation}
	\tilde{t}^{(1)\mu}(p_1,q)=P^{(1)\mu\nu}(p_1)\,q_\nu=q^\mu-c\,v_1^\mu,\qquad
	c\equiv v_1\cdot q=\frac{m_2^2-m_3^2}{2m_1}.
	\label{sm:t1-eq-q}
\end{equation}
Thus $\tilde t^{(1)\mu}=q^\mu$ only for equal daughter masses.
For unequal masses, the longitudinal term may still be
dropped when the remaining contracted spin structure is
transverse to $p_1$. This applies to the relative polarization
vectors, the tensors $\tilde\epsilon$, and transverse spinor
bilinears such as $\bar u(p_2|p_1)\gamma^\mu v(p_3|p_1)$
in Eq.~\eqref{sm:ex2-H1}.
In other cases, the cancellation must be checked in the
complete amplitude, as illustrated in Example~5.

For $L\ge2$, the trace-subtraction terms must generally be
retained, unless they vanish upon contraction with the
remaining spin structure. For example,
\[
\tilde t^{(2)\mu\nu}(p_1,q)
=\tilde t^{(1)\mu}\tilde t^{(1)\nu}
-\frac13
\big[\tilde t^{(1)}\cdot\tilde t^{(1)}\big]
\tilde g^{\mu\nu}(p_1).
\]
The first term may be replaced by $q^\mu q^\nu$ when the
remaining tensor is transverse to $p_1$ in both indices,
whereas
$\tilde t^{(1)}\cdot\tilde t^{(1)}
=q^2-(v_1\cdot q)^2=-\bm q^{*2}$.
We retain the orbital tensors below for uniformity of notation.
Moreover, with the explicit forms of the relative wave functions given in the main text one obtains, in the parent rest frame,
\begin{align}
	\varepsilon_\mu(p_2^*|k_1)=&~v^2\,\varepsilon_\mu(k_2),
	\label{sm:anchor-vec}\\
	u(p_2^*|k_1)=&~N_2\,u(k_2),\qquad
	\bar{u}(p_2^*|k_1)=N_2\,\bar{u}(k_2),\qquad
	v(p_2^*|k_1)=N_2\,v(k_2),
	\label{sm:anchor-spinor}
\end{align}
where $N_2=\sqrt{(E_2^*+m_2)/(2m_2)}$ and $k_2=(m_2,\bm{0})$, with $E_i^*$ the energy of particle $i$ in the parent rest frame, i.e., in the c.m. frame; the relative wave functions are purely spatial (or purely upper/lower components) in this frame, since $p_1\cdot\varepsilon(p_2|p_1)=0$ and $\Gamma_+v(p_2|p_1)=\Gamma_-u(p_2|p_1)=0$.
For a fermion pair the following identities hold in the parent rest frame:
\begin{align}
	\bar{u}(p_2^*|k_1)\gamma_5v(p_3^*|k_1)=&~N_2N_3\,\bar{u}(k_2)\gamma_5v(k_3),
	\label{sm:anchor-pair-0}\\
	\bar{u}(p_2^*|k_1)\gamma^0v(p_3^*|k_1)=&~0,\qquad
	\bar{u}(p_2^*|k_1)\gamma^iv(p_3^*|k_1)=N_2N_3\,\bar{u}(k_2)\gamma^iv(k_3)\quad(i=1,2,3),
	\label{sm:anchor-pair-1}
\end{align}
with $N_3=\sqrt{(E_3^*+m_3)/(2m_3)}$; i.e., in the parent rest frame the relative spinors reduce to the rest-frame spinors, up to the angle-independent factors $N_2$ and $N_3$, so that the bilinears $\bar{u}(p_2^*|k_1)\gamma_5v(p_3^*|k_1)$ and $\bar{u}(p_2^*|k_1)\bm{\gamma}v(p_3^*|k_1)$ are the pure-spin-0 and pure-spin-1 combinations of the two fermions.
Throughout this part, Lorentz indices are positioned so that every repeated (summed) index appears once as a superscript and once as a subscript.
All the simplifications below are performed with these identities, together with the transversality of the polarization vectors and of the tensors $\tilde{t}^{(L)}$ with respect to $p_1$.

\subsubsection*{Example 1: $1\to1+0$}

We consider the decay of a spin-$1$ parent into a spin-$1$ and a spin-$0$ particle ($s_1=s_2=1$, $s_3=0$), e.g., $\psi\to\rho\pi$ or $\psi\to\varphi f_0$-type processes.
From the spin quantum numbers, the allowed $L$-$S$ combinations are $S=1$ with $L=0,1,2$; the intrinsic parities select $L$ through $\eta_1=\eta_2\eta_3(-1)^{L}$ (e.g., $L=1$ for $\psi\to\rho\pi$, and $L=0,2$ for $\psi\to\varphi f_0$).
This is case~I with $\delta=s_1+s_2+s_3\bmod 2=0$, and the index-set table above gives, for the row $(s_1,s_2,s_3)=(1,1,0)$,
\begin{equation*}
	(L,S)=(0,1):\ \mathrm{I}\text{-}00,\ \#(\bm{\alpha},\bm{\beta},\bm{\gamma})=(0,0,1),\ \#(\bm{\alpha}',\bm{\beta}',\bm{\gamma}')=(1,0,0);
\end{equation*}
\begin{equation*}
	(L,S)=(1,1):\ \mathrm{I}\text{-}10,\ \#(\bm{\alpha},\bm{\beta},\bm{\gamma})=(0,0,0),\ \#(\bm{\alpha}',\bm{\beta}',\bm{\gamma}')=(1,0,0);
\end{equation*}
\begin{equation*}
	(L,S)=(2,1):\ \mathrm{I}\text{-}00,\ \#(\bm{\alpha},\bm{\beta},\bm{\gamma})=(1,1,0),\ \#(\bm{\alpha}',\bm{\beta}',\bm{\gamma}')=(1,0,0).
\end{equation*}
Substituting these index sets into the case-I formulas in Table~II of the main text (and noting that the wave function of the spin-0 particle carries no indices), one obtains
\begin{align}
	\mathcal{H}^{(0)}_{1\to1+0}=&~\varepsilon(p_1)\cdot\varepsilon^*(p_2|p_1),
	\label{sm:ex1-H0}\\
	\mathcal{H}^{(1)}_{1\to1+0}=&~i\,\varepsilon^\mu(p_1)\,q_{\nu}\,\tilde{\epsilon}^{\nu\rho}_{\mu}(p_1)\,\varepsilon^*_\rho(p_2|p_1),
	\label{sm:ex1-H1}\\
	\mathcal{H}^{(2)}_{1\to1+0}=&~\varepsilon_\alpha(p_1)\,\tilde{t}^{(2)\alpha\lambda}(p_1,q)\,\varepsilon^*_\lambda(p_2|p_1),
	\label{sm:ex1-H2}
\end{align}
where we have used Eq.~\eqref{sm:t1-eq-q} and the transversality of all the contracted objects to drop the projection tensors, and $\tilde{\epsilon}^{\nu\rho}_{\mu}(p_1)=\epsilon^{\nu\rho\alpha\beta}g_{\alpha\mu}v_{1\beta}$.
The amplitudes \eqref{sm:ex1-H0}--\eqref{sm:ex1-H2} are the complete simplified results; in the parent rest frame they reduce to
\begin{equation*}
	\mathcal{H}^{(0)}_{1\to1+0}=-v^2\,\bm{\varepsilon}(p_1)\cdot\hat{\bm{\varepsilon}},\qquad
	\mathcal{H}^{(1)}_{1\to1+0}=i\,v^2\,\bm{\varepsilon}(p_1)\cdot\left[\bm{q}^*\times\hat{\bm{\varepsilon}}\right],\qquad
	\mathcal{H}^{(2)}_{1\to1+0}=v^2\left\{\left[\bm{\varepsilon}(p_1)\cdot\bm{q}^*\right]\left[\bm{q}^*\cdot\hat{\bm{\varepsilon}}\right]-\tfrac{1}{3}\bm{q}^{*2}\,\bm{\varepsilon}(p_1)\cdot\hat{\bm{\varepsilon}}\right\},
\end{equation*}
where $\hat{\bm{\varepsilon}}$ is the rest-frame polarization of particle 2, $\varepsilon^*(p_2^*|k_1)=v^2(0,\hat{\bm{\varepsilon}})$, and the convention $\epsilon^{0123}=+1$ is used; the common factor $v^2$ stems from the normalization of the relative polarization vector, while the remaining normalizations are conventional.
These are, respectively, the pure $S$-wave ($S=1$), the standard $P$-wave, and the pure $D$-wave couplings: the angular dependence of each amplitude is carried entirely by the orbital tensor $\tilde{t}^{(L)}$, while the spin part is the pure-spin combination built from the relative wave function.

In Ref.~\cite{Zou:2002ar} the corresponding two-body vertices appear inside the chain amplitudes, e.g., for $\psi\to\varphi f_0$ the $S$- and $D$-wave vertices are given by their Eqs.~(36)--(40), and for $\psi\to\rho\pi$ the $P$-wave vertex by their Eqs.~(24)--(25): the vertex structures are $\varepsilon(p_1)\cdot\varepsilon^*(p_2)$, $\varepsilon_\alpha(p_1)\tilde{t}^{(2)\alpha\lambda}(p_1,q)\varepsilon^*_\lambda(p_2)$, and the $\epsilon$-tensor $P$-wave combination, respectively, all built from the \emph{ordinary} polarization vector $\varepsilon^*(p_2)$.
For $L=1$ our amplitude \eqref{sm:ex1-H1} and their $\epsilon$-tensor $P$-wave vertex agree up to an overall normalization, even though the two polarization vectors are not the same: the antisymmetric tensor is built from $v_1$ and $q$, both of which lie in the plane spanned by $p_1$ and $p_2$, so that it annihilates that plane, and only the component of the polarization vector orthogonal to $p_1$ and $p_2$ contributes. The ordinary and the relative polarization vectors differ by an overall factor together with a piece lying in that plane, and therefore give the same angular structure.
For $L=0$ and $L=2$, where the polarization vector is contracted directly, the forms of Ref.~\cite{Zou:2002ar} do differ from Eqs.~\eqref{sm:ex1-H0} and \eqref{sm:ex1-H2} by the replacement $\varepsilon^*(p_2)\to\varepsilon^*(p_2|p_1)$, up to the Blatt-Weisskopf factors and conventional normalization factors.
The difference can be seen explicitly in the parent rest frame.
With
$\bm{\beta}_2=\bm{p}_2^*/E_2^*$,
$\gamma_2=E_2^*/m_2$, and
$\bm{n}=\bm{p}_2^*/|\bm{p}_2^*|$,
the ordinary polarization vector is
\begin{equation*}
	\varepsilon^*(p_2^*)
	=
	\left(
	\gamma_2\bm{\beta}_2\cdot\hat{\bm{\varepsilon}},
	\;
	\hat{\bm{\varepsilon}}
	+(\gamma_2-1)\bm{n}
	(\bm{n}\cdot\hat{\bm{\varepsilon}})
	\right),
\end{equation*}
whereas the relative polarization vector is
$\varepsilon^*(p_2^*|k_1)=v^2(0,\hat{\bm{\varepsilon}})$.
Writing $\bm{\varepsilon}=\bm{\varepsilon}(k_1)$,
the ordinary vertex therefore becomes
\begin{align}
	\varepsilon(k_1)\cdot\varepsilon^*(p_2^*)
	=~
	-\bm{\varepsilon}\cdot\hat{\bm{\varepsilon}}
	-(\gamma_2-1)
	(\bm{\varepsilon}\cdot\bm{n})
	(\bm{n}\cdot\hat{\bm{\varepsilon}})
	=~
	-\frac{\gamma_2+2}{3}
	\bm{\varepsilon}\cdot\hat{\bm{\varepsilon}}
	~
	-(\gamma_2-1)
	\left[
	(\bm{\varepsilon}\cdot\bm{n})
	(\bm{n}\cdot\hat{\bm{\varepsilon}})
	-\frac{1}{3}
	\bm{\varepsilon}\cdot\hat{\bm{\varepsilon}}
	\right].
\end{align}
For fixed canonical spin states, the two terms in the final
expression have pure $S$- and $D$-wave angular structures,
respectively.
Thus the boost-dependent correction contains both an additional
$S$-wave contribution and a $D$-wave contribution.
By comparison, the relative-polarization vertex in
Eq.~\eqref{sm:ex1-H0} is proportional only to
$-\bm{\varepsilon}\cdot\hat{\bm{\varepsilon}}$,
with the angle-independent normalization factor $v^2$.

This example illustrates the distinction between the
$\mathcal{C}$-type and $\mathcal{H}$-type constructions at the
level of individual partial-wave structures.
The additional $D$-wave contribution vanishes in the
nonrelativistic limit $\gamma_2\to1$.
For the $L=1$ vertex, the antisymmetric contraction eliminates
the boost-dependent contribution, so the two constructions
have the same angular structure at any $\gamma_2$, as shown above.

\subsubsection*{Example 2: $0\to1/2+1/2$}

For the decay of a spin-$0$ particle into a fermion-antifermion pair ($s_1=0$, $s_2=s_3=1/2$), the allowed combinations are $S=0,1$ with $L=S$.
This is case~II with $\delta=1$; the index-set table above gives
\begin{equation*}
	(L,S)=(0,0):\ \mathrm{II}\text{-}01,\ \#(\bm{\alpha},\bm{\beta},\bm{\gamma})=(0,0,0),\ \#(\bm{\alpha}',\bm{\beta}',\bm{\gamma}')=(0,0,0);
\end{equation*}
\begin{equation*}
	(L,S)=(1,1):\ \mathrm{II}\text{-}00,\ \#(\bm{\alpha},\bm{\beta},\bm{\gamma})=(0,1,0),\ \#(\bm{\alpha}',\bm{\beta}',\bm{\gamma}')=(0,0,0),
\end{equation*}
and the case-II formulas in Table~II of the main text yield
\begin{align}
	\mathcal{H}^{(0)}_{0\to f\bar{f}}=&~\bar{u}(p_2|p_1)\gamma_5v(p_3|p_1),
	\label{sm:ex2-H0}\\
	\mathcal{H}^{(1)}_{0\to f\bar{f}}=&~-q_\lambda\,\bar{u}(p_2|p_1)\gamma^\lambda v(p_3|p_1).
	\label{sm:ex2-H1}
\end{align}
Here the rank-one orbital tensor can be replaced by the relative momentum: the bilinear $\bar{u}(p_2|p_1)\gamma^\lambda v(p_3|p_1)$ is transverse to $p_1$, $p_1\cdot[\bar{u}(p_2|p_1)\gamma v(p_3|p_1)]=0$ (in the parent rest frame its time component vanishes, \eqref{sm:anchor-pair-1}), so that the term $\propto v_1^\lambda$ of Eq.~\eqref{sm:t1-eq-q} drops out.
With the identities \eqref{sm:anchor-pair-0} and \eqref{sm:anchor-pair-1}, in the parent rest frame these become
\begin{equation*}
	\mathcal{H}^{(0)}_{0\to f\bar{f}}=N_2N_3\,\bar{u}(k_2)\gamma_5v(k_3),\qquad
	\mathcal{H}^{(1)}_{0\to f\bar{f}}=N_2N_3\,\bm{q}^*\cdot\bar{u}(k_2)\bm{\gamma}v(k_3),
\end{equation*}
i.e., the pure-spin-0 combination $\bar{u}\gamma_5v$ and the pure-spin-1 combination $\bar{u}\bm{\gamma}v$ of the two rest-frame spinors, with the angle-independent factors $N_2N_3$.

In Ref.~\cite{Zou:2002yy} the concept of the ``pure-spin wave functions'' of a fermion pair was introduced by hand: for the $N^*\bar{N}$ pair, the total-spin-0 and total-spin-1 wave functions are defined as $\psi^{(0)}=\bar{u}(p_B)\gamma_5v(p_C)$ and $\Psi^{(1)}_\mu=\bar{u}(p_B)(\gamma_\mu-r_\mu/(m_A+m_B+m_C))v(p_C)$ with $r=p_B-p_C$ [their Eqs.~(11)--(12)], where the $r_\mu$ term was subtracted in order to cancel the $\hat{\bm{p}}$-dependent component of $\bar{u}\gamma_\mu v$; in the parent rest frame they reduce to the pure-spin combinations $\delta$-type and $\chi^\dagger\bm{\sigma}\chi$-type [their Eqs.~(13)--(14)].
Comparing with Eqs.~\eqref{sm:ex2-H0}--\eqref{sm:ex2-H1}, the pure-spin wave functions of Ref.~\cite{Zou:2002yy} are exactly the bilinears $\bar{u}(p_2|p_1)\gamma_5v(p_3|p_1)$ and $\bar{u}(p_2|p_1)\bm{\gamma}v(p_3|p_1)$ built from the \emph{relative} spinors, up to the angle-independent normalization factors.
In other words, with the concept of the relative spin wave functions, the pure-spin wave functions of the fermion pair follow automatically and systematically: the projectors $\Gamma_\pm(p_1)$ in $u(p_2|p_1)=\Gamma_+(p_1)u(p_2)$ and $v(p_3|p_1)=\Gamma_-(p_1)v(p_3)$ play exactly the role of the $r_\mu$-subtraction of Ref.~\cite{Zou:2002yy}, and no momentum-dependent term needs to be subtracted by hand in order to obtain the pure-spin object.

\subsubsection*{Example 3: $1\to1/2+1/2$}

For the decay of a spin-$1$ parent into a fermion-antifermion pair ($s_1=1$, $s_2=s_3=1/2$), the allowed combinations are $(L,S)=(1,0)$ and $(L,S)=(0,1),(1,1),(2,1)$.
This is case~II with $\delta=0$; from the index-set table above,
\begin{equation*}
	(L,S)=(1,0):\ \mathrm{II}\text{-}01,\ \#(\bm{\alpha},\bm{\beta},\bm{\gamma})=(1,0,0),\ \#(\bm{\alpha}',\bm{\beta}',\bm{\gamma}')=(0,0,0);
\end{equation*}
\begin{equation*}
	(L,S)=(0,1):\ \mathrm{II}\text{-}00,\ \#(\bm{\alpha},\bm{\beta},\bm{\gamma})=(0,0,1),\ \#(\bm{\alpha}',\bm{\beta}',\bm{\gamma}')=(0,0,0);
\end{equation*}
\begin{equation*}
	(L,S)=(1,1):\ \mathrm{II}\text{-}10,\ \#(\bm{\alpha},\bm{\beta},\bm{\gamma})=(0,0,0),\ \#(\bm{\alpha}',\bm{\beta}',\bm{\gamma}')=(0,0,0);
\end{equation*}
\begin{equation*}
	(L,S)=(2,1):\ \mathrm{II}\text{-}00,\ \#(\bm{\alpha},\bm{\beta},\bm{\gamma})=(1,1,0),\ \#(\bm{\alpha}',\bm{\beta}',\bm{\gamma}')=(0,0,0),
\end{equation*}
and the case-II formulas in Table~II of the main text give
\begin{align}
	\mathcal{H}^{(1,0)}_{1\to f\bar{f}}=&~\varepsilon_\alpha(p_1)\,q^\alpha\,\bar{u}(p_2|p_1)\gamma_5v(p_3|p_1),
	\label{sm:ex3-H10}\\
	\mathcal{H}^{(0,1)}_{1\to f\bar{f}}=&~-\varepsilon_\lambda(p_1)\,\bar{u}(p_2|p_1)\gamma^\lambda v(p_3|p_1),
	\label{sm:ex3-H01}\\
	\mathcal{H}^{(1,1)}_{1\to f\bar{f}}=&~-i\,\varepsilon^\mu(p_1)\,q_{\nu}\,\tilde{\epsilon}^{\nu\rho}_{\mu}(p_1)\,\bar{u}(p_2|p_1)\gamma_\rho v(p_3|p_1),
	\label{sm:ex3-H11}\\
	\mathcal{H}^{(2,1)}_{1\to f\bar{f}}=&~-\varepsilon_\alpha(p_1)\,\tilde{t}^{(2)\alpha\lambda}(p_1,q)\,\bar{u}(p_2|p_1)\gamma_\lambda v(p_3|p_1).
	\label{sm:ex3-H21}
\end{align}
In the parent rest frame these become, with \eqref{sm:anchor-pair-0}--\eqref{sm:anchor-pair-1},
\begin{align*}
	\mathcal{H}^{(1,0)}_{1\to f\bar{f}}=&~-N_2N_3\left[\bm{\varepsilon}(p_1)\cdot\bm{q}^*\right]\bar{u}(k_2)\gamma_5v(k_3),\\
	\mathcal{H}^{(0,1)}_{1\to f\bar{f}}=&~N_2N_3\,\bm{\varepsilon}(p_1)\cdot\left[\bar{u}(k_2)\bm{\gamma}v(k_3)\right],\\
	\mathcal{H}^{(1,1)}_{1\to f\bar{f}}=&~-iN_2N_3\left[\bm{\varepsilon}(p_1)\times\bm{q}^*\right]\cdot\left[\bar{u}(k_2)\bm{\gamma}v(k_3)\right],\\
	\mathcal{H}^{(2,1)}_{1\to f\bar{f}}=&~N_2N_3\left\{-\left[\bm{\varepsilon}(p_1)\cdot\bm{q}^*\right]\left[\bm{q}^*\cdot\bar{u}(k_2)\bm{\gamma}v(k_3)\right]+\tfrac{1}{3}\bm{q}^{*2}\,\bm{\varepsilon}(p_1)\cdot\left[\bar{u}(k_2)\bm{\gamma}v(k_3)\right]\right\},
\end{align*}
i.e., each amplitude factorizes into the pure-spin structure of the fermion pair and an orbital factor built from $\bm{\varepsilon}(p_1)$ and $\bm{q}^*$.

For the same process, $\psi\to N^*\bar{N}$-type decays, Ref.~\cite{Zou:2002yy} lists the couplings [their Eqs.~(73)--(76)]: $(S_{BC},L_{BC})=(1,0)$: $\Psi^{(1)}_\mu\varepsilon^\mu$; $(1,2)$: $\Psi^{(1)}_\mu\varepsilon_\nu\tilde{t}^{(2)\mu\nu}$; $(0,1)$: $\psi^{(0)}\varepsilon_\mu\tilde{t}^{(1)\mu}$; $(1,1)$: $i\Psi^{(1)}_\mu\epsilon^{\mu\nu\lambda\sigma}\varepsilon_\nu\tilde{t}^{(1)}_\lambda\hat{p}_{(\psi)\sigma}$, where $\psi^{(0)}$ and $\Psi^{(1)}_\mu$ are their pure-spin wave functions of Eqs.~(11)--(12), $\varepsilon^\mu$ is the polarization vector of the $\psi$, and $\hat{p}_{(\psi)}=p_\psi/m_\psi$.
Comparing with Eqs.~\eqref{sm:ex3-H10}--\eqref{sm:ex3-H21} term by term, the two results are indeed consistent: each of our amplitudes has the same spin and orbital structure as the corresponding coupling of Ref.~\cite{Zou:2002yy}, with the relative spinors $\bar{u}(p_2|p_1)\gamma_5v(p_3|p_1)$ and $\bar{u}(p_2|p_1)\gamma^\lambda v(p_3|p_1)$ playing the role of their $\psi^{(0)}$ and $\Psi^{(1)}_\mu$, and with their $\hat{p}_{(\psi)}$ and $r_\mu$-subtraction replaced by the covariant $\tilde{\epsilon}$-tensor built from $v_1$ and by the projectors $\Gamma_\pm$, respectively; the remaining differences are conventional normalization factors.

\subsubsection*{Example 4: $1\to1/2+3/2$}

For the decay of a spin-$1$ parent into a spin-$1/2$ and a spin-$3/2$ fermion ($s_1=1$, $s_2=1/2$, $s_3=3/2$), the allowed combinations are $S=1$ with $L=0,1,2$ and $S=2$ with $L=1,2,3$.
This is case~II with $\delta=1$; the index-set table above gives, for the parity-allowed cases of $\psi\to N^*(3/2^+)\bar{N}$-type decays ($L=0,2$ with $S=1$ and $L=2$ with $S=2$),
\begin{equation*}
	(L,S)=(0,1):\ \mathrm{II}\text{-}01,\ \#(\bm{\alpha},\bm{\beta},\bm{\gamma})=(0,0,1),\ \#(\bm{\alpha}',\bm{\beta}',\bm{\gamma}')=(0,0,1);
\end{equation*}
\begin{equation*}
	(L,S)=(2,1):\ \mathrm{II}\text{-}01,\ \#(\bm{\alpha},\bm{\beta},\bm{\gamma})=(1,1,0),\ \#(\bm{\alpha}',\bm{\beta}',\bm{\gamma}')=(0,0,1);
\end{equation*}
\begin{equation*}
	(L,S)=(2,2):\ \mathrm{II}\text{-}10,\ \#(\bm{\alpha},\bm{\beta},\bm{\gamma})=(0,1,0),\ \#(\bm{\alpha}',\bm{\beta}',\bm{\gamma}')=(0,0,1),
\end{equation*}
and the case-II formulas in Table~II of the main text yield (after dropping $P^{(1)}(p_1)$ by transversality)
\begin{align}
	\mathcal{H}^{(0,1)}_{1\to f_{1/2}f_{3/2}}=&~\varepsilon_\rho(p_1)\,\bar{u}(p_2|p_1)\gamma_5v^{\rho}(p_3|p_1),
	\label{sm:ex4-H01}\\
	\mathcal{H}^{(2,1)}_{1\to f_{1/2}f_{3/2}}=&~\varepsilon_\alpha(p_1)\,\tilde{t}^{(2)\alpha\rho}(p_1,q)\,\bar{u}(p_2|p_1)\gamma_5v_{\rho}(p_3|p_1),
	\label{sm:ex4-H21}\\
	\mathcal{H}^{(2,2)}_{1\to f_{1/2}f_{3/2}}=&~-i\,\varepsilon^\mu(p_1)\,\tilde{t}^{(2)\beta}_{\nu}(p_1,q)\,\tilde{\epsilon}^{\nu\rho}_{\mu}(p_1)\,P^{(2)}_{\beta\rho;\kappa\lambda}(p_1)\,\bar{u}(p_2|p_1)\gamma^{\lambda}v^{\kappa}(p_3|p_1),
	\label{sm:ex4-H22}
\end{align}
where the spin-$3/2$ projection tensor $P^{(3/2)}(p_1)$ from Eq.~(21) of the main text acts as the identity on the relative vector-spinor $v^\rho(p_3|p_1)$ because the latter already satisfies the anchor's Rarita-Schwinger constraints $p_1^\rho v_\rho(p_3|p_1)=0$ and $\gamma^\rho v_\rho(p_3|p_1)=0$, and reduces to $v_\rho(p_3^*|k_1)=N_3^3\,v_\rho(k_3)$ in the parent rest frame.
The simplification in the parent rest frame proceeds as in the previous examples: the pure-spin parts $\bar{u}(p_2|p_1)\gamma_5v_\rho(p_3|p_1)$ and $P^{(2)}_{\beta\rho;\kappa\lambda}(p_1)\bar{u}(p_2|p_1)\gamma^\lambda v^\kappa(p_3|p_1)$ reduce to the rest-frame spin-$1$ and traceless spin-$2$ bilinears, respectively, with the spin-$2$ projection tensor $P^{(2)}$ in Eq.~\eqref{sm:ex4-H22} projecting the fermion pair onto definite total spin $S=2$.

For the same process Ref.~\cite{Zou:2002yy} lists, for $\psi\to N^*(3/2^+)\bar{N}$ [their Eqs.~(77)--(79)], the couplings $(S_{BC},L_{BC})=(1,0)$: $\psi^{(1)}_\mu\varepsilon^\mu$; $(1,2)$: $\psi^{(1)}_\mu\varepsilon_\nu\tilde{t}^{(2)\mu\nu}$; $(2,2)$: $i\Psi^{(2)}_{\mu\alpha}\epsilon^{\mu\nu\lambda\sigma}\varepsilon_\nu\tilde{t}^{(2)\alpha}_{\lambda}\hat{p}_{(\psi)\sigma}$, with $\psi^{(1)}_\mu=\bar{u}_\mu(p_B)\gamma_5v(p_C)$ and $\Psi^{(2)}_{\mu\nu}=\bar{u}_\mu(p_B)(\gamma_\nu-r_\nu/(m_A+m_B+m_C))v(p_C)+(\mu\leftrightarrow\nu)$ built from the \emph{ordinary} Rarita-Schwinger wave function $u_\mu(p_B)$ of the spin-$3/2$ fermion.
Here the two constructions \emph{do not} agree.
Both $\psi^{(1)}_\mu$ and $\Psi^{(2)}_{\mu\nu}$
[Eqs.~(9) and (10) of Ref.~\cite{Zou:2002yy}, with $n=1$]
are built from the ordinary boosted vector-spinor $u_\mu(p_B)$.
Its vector index retains $\hat{\bm{p}}$ dependence in the
parent rest frame, which can generate mixing between canonical
partial waves.
In the RSWF construction, the vector and spinor indices are
transformed by $\Pi$ and $\Gamma_\pm$, respectively.
Their combined action yields a vector-spinor proportional
to its rest-frame counterpart in the parent rest frame.

\subsubsection*{Example 5: $1/2\to1+1/2$}

For the decay of a spin-$1/2$ parent into a spin-$1$ and a spin-$1/2$ particle ($s_1=1/2$, $s_2=1$, $s_3=1/2$), e.g., $N^*\to N\omega$-type processes, the allowed combinations are $S=1/2$ with $L=0,1$ and $S=3/2$ with $L=1,2$.
This is case~III with $\delta=0$; the index-set table above gives
\begin{equation*}
	(L,S)=(0,\tfrac12):\ \mathrm{III}\text{-}10,\ \#(\bm{\alpha},\bm{\beta},\bm{\gamma})=(0,0,0),\ \#(\bm{\alpha}',\bm{\beta}',\bm{\gamma}')=(0,0,0);
\end{equation*}
\begin{equation*}
	(L,S)=(1,\tfrac12):\ \mathrm{III}\text{-}00,\ \#(\bm{\alpha},\bm{\beta},\bm{\gamma})=(0,0,0),\ \#(\bm{\alpha}',\bm{\beta}',\bm{\gamma}')=(0,0,0);
\end{equation*}
\begin{equation*}
	(L,S)=(1,\tfrac32):\ \mathrm{III}\text{-}11,\ \#(\bm{\alpha},\bm{\beta},\bm{\gamma})=(0,1,0),\ \#(\bm{\alpha}',\bm{\beta}',\bm{\gamma}')=(1,0,0);
\end{equation*}
\begin{equation*}
	(L,S)=(2,\tfrac32):\ \mathrm{III}\text{-}01,\ \#(\bm{\alpha},\bm{\beta},\bm{\gamma})=(0,1,0),\ \#(\bm{\alpha}',\bm{\beta}',\bm{\gamma}')=(1,0,0),
\end{equation*}
and the case-III formulas in Table~II of the main text yield
\begin{align}
	\mathcal{H}^{(0,\frac12)}_{1/2\to1+1/2}=&~\varepsilon^*_\mu(p_2|p_1)\,\bar{u}(p_3|p_1)\gamma_5\gamma^\mu u(p_1),
	\label{sm:ex5-H0}\\
	\mathcal{H}^{(1,\frac12)}_{1/2\to1+1/2}=&~\varepsilon^*_\mu(p_2|p_1)\,\tilde{t}^{(1)}_{\nu}(p_1,q)\,\bar{u}(p_3|p_1)\gamma^\mu\gamma^\nu u(p_1),
	\label{sm:ex5-H1}\\
	\mathcal{H}^{(1,\frac32)}_{1/2\to1+1/2}=&~-\varepsilon^{*\alpha}(p_2|p_1)\,\tilde{t}^{(1)\beta}(p_1,q)\,\bar{u}(p_3|p_1)P^{(3/2)}_{\alpha\beta}(p_1)u(p_1),
	\label{sm:ex5-H13}\\
	\mathcal{H}^{(2,\frac32)}_{1/2\to1+1/2}=&~\varepsilon^{*\alpha}(p_2|p_1)\,\tilde{t}^{(2)\beta}_{\mu}(p_1,q)\,\bar{u}(p_3|p_1)P^{(3/2)}_{\alpha\beta}(p_1)\gamma_5\gamma^\mu u(p_1),
	\label{sm:ex5-H23}
\end{align}
where the spin-$1/2$ projection tensor has been dropped since $P^{(1/2)}$ acts as the identity on the spinor indices.
In the parent rest frame, with $\varepsilon^*(p_2|p_1)=v^2(0,\hat{\bm{\varepsilon}})$ and $\bar{u}(p_3|p_1)=N_3\bar{u}(k_3)$, the first amplitude becomes
\begin{equation*}
	\mathcal{H}^{(0,\frac12)}_{1/2\to1+1/2}=-v^2N_3\,\hat{\bm{\varepsilon}}\cdot\left[\bar{u}(k_3)\gamma_5\bm{\gamma}u(p_1)\right],
\end{equation*}
i.e., the pure-spin-$1/2$ coupling of the vector meson and the nucleon, with no further angular-momentum recoupling.
In the two $L=1$ amplitudes, the longitudinal subtraction in
$\tilde{t}^{(1)}$ vanishes after the complete spin-polarization
contraction.
For Eq.~\eqref{sm:ex5-H1}, using
$\Gamma_+(p_1)\gamma^\mu\Gamma_+(p_1)
=v_1^\mu\Gamma_+(p_1)$ and the anchor transversality of the
relative polarization vector, we obtain
\begin{equation*}
	p_{1\nu}\varepsilon_\mu^*(p_2|p_1)
	\bar{u}(p_3|p_1)\gamma^\mu\gamma^\nu u(p_1)
	=
	m_1\big[\varepsilon^*(p_2|p_1)\cdot v_1\big]
	\bar{u}(p_3|p_1)u(p_1)=0.
\end{equation*}
For Eq.~\eqref{sm:ex5-H13}, the same conclusion follows from
$p_1^\beta P^{(3/2)}_{\alpha\beta}(p_1)=0$.
Thus $\tilde{t}^{(1)}$ may equivalently be replaced by $q$
in these two amplitudes, including for unequal daughter masses.
We retain the orbital tensors in the expressions above
for uniformity of notation.

Reference~\cite{Zou:2002yy} organizes the spin structures
through fermion-pair tensors:
$\psi^{(n)}$ and $\Psi^{(n+1)}$ for a fermion-antifermion
pair [their Eqs.~(9)--(10)], and
$\phi^{(n)}$ and $\Phi^{(n+1)}$ for the parent and daughter
fermions [their Eqs.~(17)--(22)].
Therefore, when the formulas of Ref.~\cite{Zou:2002yy} are used to construct the partial-wave amplitudes of processes with a fermion pair involved, the spin angular momentum $S$ in their couplings always refers to the total spin of a \emph{fermion pair}: in their notation the couplings are labelled by $(S_A,S_{BC},L_{BC})$ for a mesonic parent and by $(S_C,S_{AB},L_{BC})$ for a fermionic parent, where $S_{BC}$ and $S_{AB}$ are the spins of the fermion pair.
For case-III processes such as $N^*\to N\omega$, however, the total spin $S$ of the two final particles is the total spin of a fermion and a \emph{boson}, not that of a fermion pair; in Ref.~\cite{Zou:2002yy} one therefore has to first couple the two fermions (the parent and the final nucleon) into $S_{AB}$ with $\phi^{(n)}$ or $\Phi^{(n+1)}$, and then recouple $S_{AB}$ with the spin of the boson to obtain the final-state spin $S$, which requires additional angular-momentum recouplings with Wigner-$6j$ coefficients.
For example, their amplitude for $N^*(3/2^-)\to N\omega$ involves three couplings $g_1\varphi^{(1)}_\mu\varepsilon^{*\mu}+g_2\varphi^{(1)}_\mu\varepsilon^*_\nu\tilde{t}^{(2)\mu\nu}+g_3i\Phi^{(2)}_{\mu\alpha}\epsilon^{\mu\nu\lambda\sigma}\varepsilon^*_\nu\tilde{t}^{(2)\alpha}_{\lambda}\hat{p}^*_\sigma$ [their Eq.~(64)], where $\varepsilon^{*\mu}$ is the polarization vector of the $\omega$ and $\varphi^{(1)}_\mu$, $\Phi^{(2)}_{\mu\alpha}$ are the fermion-pair objects of their Eqs.~(17)--(22); expressing the final state in the conventional $(L,S)$ scheme in which $S$ is the total spin of the $N\omega$ pair requires the $6j$ recoupling of $(S_{AB},S_\omega)$ into $S$, which considerably complicates the use of the formulas of Ref.~\cite{Zou:2002yy}.
The construction of this work does not suffer from this complication: the total spin $S$ of the two final particles is built directly into the coupling tensor $P^{(S)}_{\bm{\alpha}'\bm{\gamma}';\bm{\beta}\bm{\gamma}}(p_1)$ [see Eqs.~\eqref{sm:ex5-H0}--\eqref{sm:ex5-H23}], so that no further recoupling is needed to define $S$ as the total spin of the two final particles at this vertex.

	\twocolumngrid
	
\end{document}